\documentclass[aps,pra,twocolumn,byrevtex,showpacs,floatfix,superscriptaddress,reprint]{revtex4-2}%,reprint
\usepackage{graphicx}
\usepackage{subcaption}
\usepackage{amsmath,amsfonts,amsthm,bm}\usepackage{epsfig}
\usepackage{hyperref}

\usepackage{caption}
\graphicspath{{pict/}{}}

\usepackage{mathrsfs}
\usepackage{mathtools}

\usepackage{algorithm} 
\usepackage[noend]{algpseudocode}

\newcommand{\lp}[1]{\textcolor{black}{#1}}

\hypersetup{colorlinks=true,linkcolor=blue,citecolor=blue,urlcolor=blue}

\begin{document}

\title{Detecting entanglement of unknown states by violating the Clauser-Horne-Shimony-Holt inequality}

\author{J. Cort\'es-Vega}
\email[corresponding author: ]{jeancortes@udec.cl}
\affiliation{Instituto Milenio de Investigaci\'on en \'Optica, Universidad de Concepci\'on, Concepci\'on, Chile}
\affiliation{Facultad de Ciencias F\'isicas y Matem\'aticas, Departamento de F\'isica, Universidad de Concepci\'on, Concepci\'on, Chile}

\author{J. F. Barra}
\affiliation{Instituto Milenio de Investigaci\'on en \'Optica, Universidad de Concepci\'on, Concepci\'on, Chile}
\affiliation{Facultad de Ciencias F\'isicas y Matem\'aticas, Departamento de F\'isica, Universidad de Concepci\'on, Concepci\'on, Chile}

\author{L. Pereira}
\affiliation{Instituto de F\'{\i}sica Fundamental IFF-CSIC, Calle Serrano 113b, Madrid 28006, Spain}

\author{A. Delgado}
\affiliation{Instituto Milenio de Investigaci\'on en \'Optica, Universidad de Concepci\'on, Concepci\'on, Chile}
\affiliation{Facultad de Ciencias F\'isicas y Matem\'aticas, Departamento de F\'isica, Universidad de Concepci\'on, Concepci\'on, Chile}

\date{\today}

%\pacs{42.50.Dv, 03.67.-a}
% 42.50.Dv Quantum state engineering and measurements  
% 03.67.-a Quantum information 
% 03.65.-w Quantum mechanics
% 03.67.Hk Quantum communication

\begin{abstract}

Entangled states play a fundamental role in Quantum Mechanics and are at the core of many contemporary applications, such as quantum communication and quantum computing. Therefore, determining whether a state is entangled or not is an important task. Here, we propose a method to detect the entanglement of unknown two-qubit quantum states. Our method is based on the violation of the Clauser-Horne-Shimony-Holt inequality. This maximizes the value of the inequality even when \lp{it} contains an unknown quantum state. The method iteratively generates local measurement settings that lead to increasing values of the inequality. \lp{We show by numerical simulations for pure and mixed states that our algorithm exceeds the classical limit of 2 after a few iterations. }

\end{abstract}

\maketitle

\section{Introduction}

Quantum mechanics predicts the existence of quantum states of composite systems that cannot be written
as products of states of their individual components \cite{von_Neumann}. These are the so called entangled states. Today, these states play a central role in quantum information theory \cite{Alber,Horodecki1} and in many applications, such as, for instance, quantum cryptography \cite{Ekert}, quantum teleportation \cite{Bennett,Boschi1998}, frequency standards improvement \cite{Wineland,Huelga,Giovannetti}, one-way quantum computing \cite{Raussendorf}, clock synchronization \cite{Jozsa}, and entanglement assisted orientation in space \cite{Bruckner}, among many others. Interestingly, entangled states play a key role in the argument put forward by Einstein, Podolsky, and Rosen \cite{Einstein}. This was aimed at ascribing objective values to measurable quantities, that is, values that exist prior to and independently of measurements. Bell's inequality \cite{Bell} shows that precisely the existence of entangled states precludes such conception of reality. 

In view of the foundational significance of entangled states and their many applications, theoretical and experimental characterization and detection of entangled states are important research subjects. One of the first criteria employed to study the entanglement of quantum states is the violation of the Clauser-Horne-Shimony-Holt inequality \cite{CHSH,Aspect} (CHSH), which is the  generalization of Bell's inequality to two observers each having the choice of two measurement settings with two outcomes. In this scenario, the violation of the CHSH inequality indicates the presence of entanglement. This approach has also been studied in the context of the theory of entanglement witnesses \cite{Horodecki,Terhal}. These are observables with positive expectation values with respect to the complete set of separable states that for at least one entangled state provide a negative expectation value. Thus, a negative expectation value signals the presence of entanglement. It has been shown that the CHSH inequality can be related to an entanglement witness \cite{Terhal,Hyllus}.

Here, we study the detection of entanglement of unknown states via the violation of the CHSH inequality. 
Since the majority of the entanglement measures and entanglement detectors are based on the knowledge of the quantum state, the unknown character of the state increases the difficulty of the problem. \lp{The presence of unknown quantum states is common in quantum communication \cite{QComm1,QComm2, QComm3} and quantum computing \cite{QComp1,QComp2,QComp3}, where an objective entangled state is prepared, but it is modified by the action of the environment.} Entanglement detection of unknown quantum states has been previously studied from the point of view of quantum tomography \cite{DAriano,Paris} by means of an adaptive scheme \cite{Maciel,Lima}, employing a succession of measurements of witness operators \cite{Zhu,Dai}, via the measurement of the energy observable \cite{Cavalcanti}, via local parity measurements on two-fold copies of the unknown state \cite{Mintert}, series of local random measurements from which entanglement witnesses are constructed \cite{Szangolies}, and variational determination of geometrical entanglement \cite{VDGE}, among many others. We follow a different approach. For a given known state, the maximal violation of the CHSH inequality is obtained by maximizing the inequality onto the set of 4-tuples of dichotomic observables. This procedure is typically carried out by means of semidefinite programming (SDP) techniques. If the state is unknown, then the function to be optimized, that is, the target function, contains unknown fix parameters and SDP cannot be employed to find the measurements leading to the maximal violation. Analogously, the use of an entanglement witness also requires the knowledge about the state. To overcome this problem we employ a recently developed optimization algorithm \cite{Utreras-Alarcon}, the Complex simultaneous perturbation stochastic approximation (CSPSA), which can handle functions with unknown parameters. CSPSA works natively within the field of the complex numbers. Thereby, no parameterization of the complex arguments onto the real numbers is necessary. Also, this algorithm has exhibited an improved convergence rate in certain applications such as, for instance, the estimation of unknown quantum pure states \cite{Zambrano}. CSPSA uses a stochastic approximation of the complex Wirtinger gradient of the target function, that is, the function to be optimized, which requires the value of the target function at two different points in the optimization space. In the case at hand, these two values can be obtained experimentally, regardless of whether the state remains unknown. CSPSA iteratively generates a sequence of sets with four local measurement settings with increasing values of the CHSH function until reaching the highest possible violation of the inequality.  

We first study via numerical simulations the performance of the method here proposed when applied to unknown pure 2-qubit states. In this case, the maximal value achieved by the CHSH function depends on the Schmidt coefficient of the state. Thereby, the performance of the method can be compared with an analytical bound.  We show that for the set formed by states that have the same set of local Schmidt bases, the method leads in tens of iterations to a value close to the maximum of the CHSH function for each value of the Schmidt coefficient. We also consider sets of states that have the same concurrence value but different local Schmidt bases. In this case, the method also approaches the corresponding maximum value of the CHSH inequality in tens of iterations. However, the higher the concurrence value, the fewer iterations are required for a violation of the CHSH inequality. Also, all states with the same concurrence value exhibit a very similar behavior of the CHSH function as a function of the number of iterations, that is, CSPSA produces results that are nearly independent of the particular set of local Schmidt bases. We also consider the average behavior of the method on the Hilbert space of two qubits. In this case, the method reaches a CHSH function value greater than 2 after 17 iterations for an ensemble size of $10^2$. After 25 iterations the interquartile range is also above 2, which indicates that for 75\% of the simulated states the method reached a violation of the CHSH inequality. A further increase of the ensemble size leads to a reduction in the number of iterations required to achieve a violation of the CHSH inequality. In order to study the accuracy achieved by our method we employ the squared error. We show that the mean and median squared error on the 2-qubit Hilbert space are nearly indistinguishable. After 25 iterations the mean square error achieves a value in the order of $10^{-1}$ for an ensemble size of $10^2$. A further increase of the ensemble size to $10^3$ leads to a decrease in mean square error in the order of half order of magnitude. Thereafter, we study the case of two-qubit mixed states. Unlike the case of pure states, there is no known analytical formula for the maximum value of the CHSH function for an arbitrary mixed state. However, in the particular case of Werner states, that is, a maximally entangled state affected by white noise, it is possible to obtain the maximum value of the CHSH function in terms of the mixing parameter. We show that CSPSA is capable of achieving a value close to the maximum violation of the CHSH inequality for all Werner states. As the ensemble size increases the value of the function provided by CSPSA becomes closer to the maximal violation. Finally, we analyze the results achieved by CSPSA for unknown mixed states. \lp{For these states there is no analytical expression for maximal violation, so we calculate this value via SDP.} After generating $10^6$ density matrices, a subset of $8\times10^3$ density matrices that violate the CHSH inequality is identified. These states have a small value of the negativity, a well-known entanglement measure. Within this subset, the mean and median values of the CHSH function provided by CSPSA achieve a value close to the theoretical maximal violation after approximately 75 iterations. 

Our results show that the maximization of the CHSH function via the CSPSA method allows detecting the entanglement of unknown states, pure or mixed, with a high degree of accuracy. Furthermore, the highest value of the CHSH function can also be achieved. Our approach requires the ability to adapt local measurements, which are carried out on single copies of the unknown state. This can be implemented in various experimental platforms \cite{Aspect,Hensen,Carvacho,Giustina,Abellan,Vedovato}. We stress the fact that no {\it a priori} information about the unknown state, such as purity, Schmidt coefficient, or Schmidt bases, has been employed to optimize the performance of CSPSA. 

\section{CHSH inequality and CSPSA optimization algorithm} 

The target function to be optimized is the Clauser-Horne-Shimony-Holt function $S$ defined by the expression \cite{CHSH}
\begin{eqnarray}
S({\bm z},{\bm z}^*)&=&E({\bm z}_a,{\bm z}_b)+E({\bm z}_a,{\bm z}'_b)+E({\bm z}'_a, {\bm z}_b)
\nonumber\\
&-&E({\bm z}'_a,{\bm z}'_b),
\end{eqnarray}
where the expectation value  $E({\bm z}_a,{\bm z}_b)$ is given by the average of the products of the outcomes of two locally performed dichotomic measurements $A({\bm z}_a)$ and $B({\bm z}_b)$ defined by the settings ${\bm z}_a$ and ${\bm z}_b$, respectively. The vector ${\bm z}$ contains the settings of the four local measurements, that is, $\bm z=({\bm z}_a, {\bm z}'_a, {\bm z}_b, {\bm z}'_b)$. The CHSH inequality adopts the form $|S|\le2$. 

A quantum mechanical dichotomic observable $A({\bm z}_a)$ is defined as the one having $\pm1$ eigenvalues, that is, an observable with the spectral decomposition
\begin{equation}
A({\bm z}_a)=|\psi({\bm z}_a)\rangle\langle \psi({\bm z}_a)|-|\psi^\perp({\bm z}_a)\rangle\langle \psi^\perp({\bm z}_a)|,
\label{LocalObservable}
\end{equation}
where $|\psi({\bm z}_a)\rangle$ is an arbitrary two-dimensional quantum state
\begin{equation}
|\psi({\bm z}_a)\rangle=\frac{z_{a,1}|0\rangle+z_{a,2}|1\rangle}{\sqrt{|z_{a,1}|^2+|z_{a,2}|^2}}. 
\label{LocalObservableState}
\end{equation}
The state $|\psi^\perp({\bm z}_a)\rangle$ is orthogonal to $|\psi({\bm z}_a)\rangle$ and the components $z_{a,1}$ and $z_{a,2}$ of the vector ${\bm z}_a$ are complex numbers. Thereby, the expectation value $E({\bm z}_a,{\bm z}_b)$ is given by the expression
\begin{eqnarray}
E({\bm z}_a,{\bm z}_b)&=&Tr(\rho|\psi({\bm z}_a)\rangle\langle \psi({\bm z}_a)|\otimes|\psi({\bm z}_b)\rangle\langle \psi({\bm z}_b)|)
\nonumber\\
&+&Tr(\rho|\psi^\perp({\bm z}_a)\rangle\langle \psi^\perp({\bm z}_a)|\otimes|\psi^\perp({\bm z}_b)\rangle\langle \psi^\perp({\bm z}_b)|)
\nonumber\\
&-&Tr(\rho|\psi({\bm z}_a)\rangle\langle \psi({\bm z}_a)|\otimes|\psi^\perp({\bm z}_b)\rangle\langle \psi^\perp({\bm z}_b)|)
\nonumber\\
&-&Tr(\rho|\psi^\perp({\bm z}_a)\rangle\langle \psi^\perp({\bm z}_a)|\otimes|\psi({\bm z}_b)\rangle\langle \psi({\bm z}_b)|),
\end{eqnarray}
where $\rho$ is a fixed known two-qubit state. 

The problem of violating the CHSH inequality consists in finding a complex vector ${\bm z}$ such that for a given known state $\rho$ leads to a maximal value of $|S({\bm z},{\bm z}^*)|$ larger than the classical bound of 2. This optimization problem can be solved by means of semidefinite programing or other numerical optimization techniques. However, when the state $\rho$ entering in the function $S$ is unknown, the standard approaches to the problem cannot be employed. The reason for this is that the function $S$ and its derivatives cannot be evaluated. 

\begin{algorithm}[H]
\caption{CSPSA optimization of S($\rho, {\bm z}$)} 
\label{Algorithm1}
\begin{algorithmic}
\State Consider a state $\rho$. This plays the role of the unknown state.
\State Set initial guess $\hat{{\bm z}}_0$, and gain coefficients $a$, $A$, $s$, $b$, and $r$.
\For {$k=1,\dots, k_{max}$}
\State Set $$a_k =\frac{a}{(k+1+A)^s},\quad c_k = \frac{b}{(k +1)^r}.$$
\State Choose $\Delta_{k,i}$ randomly in the set $\{\pm1,\pm i \}$.
\State Calculate $\hat{{\bm z}}_{k\pm}=\hat{{\bm z}}_k\pm c_k{\bm \Delta_k}$.
\State Estimate from experimentally acquired data or numerically simulate the values $S(\rho, \hat{{\bm z}}_{k\pm})$ considering an ensemble of $N$ equally prepared pairs of qubits in the state $\rho$.
\State Estimate the gradient as $$\hat{g}_{k,i} = \frac{S(\rho, \hat{{\bm z}}_{k+})-S(\rho, \hat{{\bm z}}_{k-})}{2c_k\Delta_{k,i}^*}.$$
\State Actualize the guess $\hat{{\bm z}}_{k+1}=\hat{{\bm z}}_k+a_k \hat{{\bm g}}_k$.
\State Normalize coefficients $\hat{{\bm z}}_{k+1}$
\EndFor
\end{algorithmic} 

\end{algorithm}

In order to overcome this problem, we resort to the recently introduced CSPSA \cite{Utreras-Alarcon} optimization algorithm for real-valued functions of complex arguments. This algorithm works natively on the field of the complex numbers, which make unnecessary the use of real parameterizations of the complex arguments. For a target function  $f({\bm z},{\bm z}^*):\mathbb{C}^n\times\mathbb{C}^n\rightarrow\mathbb{R}$, CSPSA is defined by the iterative rule
\begin{equation}
\hat{\bm z}_{k+1}=\hat{\bm z}_k+a_k\hat{\bm g}_k(\hat{\bm z}_k,\hat{\bm z}_k^*),
\label{ALGORITHM}
\end{equation} 
where $a_k$ is a positive gain coefficient and $\hat{\bm z}_k$ is the estimate of the maximizer $\tilde{\bm z}$ of $f({\bm z},{\bm z}^*)$ at the k-th iteration. The iteration starts from an initial guess $\hat{\bm z}_0$, which is randomly chosen. The function $\hat{\bm g}_k(\hat{\bm z}_k,\hat{\bm z}_k^*)$ is an estimator for the Wirtinger gradient \cite{Wirtinger} of $f({\bm z},{\bm z}^*)$ whose components are defined by
\begin{equation}
\hat{g}_{k,i}=\frac{f(\hat{\bm z}_{k+},\hat{\bm z}_{k+}^*)+\epsilon_{k,+}-(f(\hat{\bm z}_{k-},\hat{\bm z}_{k-}^*)+\epsilon_{k,-})}{2c_k{\Delta}_{k,i}^*},
\label{ESTGRADIENT}
\end{equation}
with 
\begin{equation}
\hat{\bm z}_{k\pm}=\hat{\bm z}_k\pm c_k{\bm\Delta}_k,
\label{zetasplusminus}
\end{equation}
where $c_k$ is a positive gain coefficient and $\epsilon_{k,\pm}$ describes the presence of noise in the values of $f(\hat{\bm z}_{k\pm},\hat{\bm z}_{k\pm}^*)$. The components of the vector ${\bm \Delta}_k\in\mathbb{C}^n$ are identically and independently distributed random variables in the set $\{\pm1,\pm i\}$. The gain coefficients $a_k$ and $c_k$ control the convergence of CSPSA and are chosen as
\begin{equation}
a_k=\frac{a}{(k+1+A)^s},~~c_k=\frac{b}{(k+1)^r}.
\end{equation}
The values of $a, A, s, b$ and $r$ are adjusted to optimize the rate of convergence depending on the target function. We use the values: $a = 1.0$, $b = 0.25$, $s = 1.0$, $r = 1/6$, and $A = 0$. 

Two main properties of CSPSA are: (i) it converges asymptotically in mean to the maximizer $\tilde{\bm z}$ of $f({\bm z},{\bm z}^*)$ and (ii) $\hat{\bm g}_k$ is an asymptotically unbiased estimator of the Wirtinger gradient. With proper conditions, these properties are maintained even in the presence of the noise terms $\epsilon_{k,\pm}$ entering in Eq.\thinspace(\ref{ESTGRADIENT}). CSPSA is the generalization of the Simultaneous perturbation stochastic approach (SPSA) \cite{Spall1,Spall2} from the field of real numbers to the field of complex numbers. SPSA has been applied to the problem of estimating pure states \cite{Ferrie,Granade} and experimentally realized \cite{Chapman}. 

Thus, the application of CSPSA to the maximization of the CHSH function proceeds as follows: an initial guess $\hat{\bm z}_0$ for the vector containing the measurement settings and a vector ${\bm\Delta}_0$ are randomly generated. These two vectors are employed to calculate the vectors $\hat{\bm z}_{0\pm}$ according to Eq.~(\ref{zetasplusminus}). Thereafter, the values $S(\hat{\bm z}_{0\pm},\hat{\bm z}^*_{0\pm})$ of the CHSH function are obtained, which involves the realization of measurements on a finite ensemble of $N$ copies of the unknown state $\rho$. The values $S(\hat{\bm z}_{0\pm},\hat{\bm z}^*_{0\pm})$ are then employed to calculate the estimator for the Wirtinger gradient $\hat{\bm g}_0(\hat{\bm z}_0,\hat{\bm z}_0^*)$ using Eq.~(\ref{ESTGRADIENT}). Finally,  a new estimate $\hat{\bm z}_1$ for the vector of settings is obtained by means of Eq.~(\ref{ALGORITHM}). This process is iterated until achieving a violation of the CHSH inequality or until reaching a predefined number of iterations. Algorithm \ref{Algorithm1} shows a pseudocode for the optimization of the CHSH function via CSPSA.

According to Eq.\thinspace(\ref{ESTGRADIENT}), the use of CSPSA to maximize the CHSH function $S({\bm z},{\bm z}^*)$ requires the capability of obtaining the values $S(\hat{\bm z}_{k\pm},\hat{\bm z}_{k\pm}^*)$ at each iteration, which in turn requires an experimental platform capable of measuring the CHSH function at any value of the setting vector ${\bm z}$. In photonic platforms where a qubit is encoded in the polarization degree of freedom of a single photon, the local measurements on a qubit are carried out by the interaction of the photon with a sequence of half- and quarter-wave plates followed by a polarizing beam splitter and single-photon detectors. In this case a setting vector is given by the rotation angles of the wave plates. Thereby, it is possible to implement any local measurement up to the angular resolution of the wave plates. It is possible to achieve a high degree of control in other experimental platforms, for instance in time-bin or energy-time encoded qubits, where local measurements can be implemented introducing electronically controlled phase shifts. Thus, we will assume that the CHSH function can be measured for any value of the setting vector ${\bm z}$.

\section{Results}

A single run of CSPSA starts with the choice of an initial guess ${\bm z}_0$ of the four local measurement bases and proceeds through the choice of the vector ${\bm\Delta}_k$ at every iteration. Since there is no a priori information about the initial state, the initial guess for each of the local measurements, which are defined by Eqs.~(\ref{LocalObservable}) and (\ref{LocalObservableState}), is randomly chosen according to a Haar-uniform distribution. The choice of ${\bm\Delta}_k$ is equally random. Thereby, CSPSA is an intrinsically stochastic optimization algorithm. A third source of randomness is the value of the CHSH function. This is obtained  by means of probabilities that are inferred from local measurements made on a set of equally prepared copies of the unknown state. Since the size $N$ of the ensemble is finite, the inferred probabilities are affected by finite statistic noise. Thereby, CSPSA exhibits three different sources of randomness and, consequently, each run of CSPSA will follow a different trajectory in the optimization space, that is, the space of all four setting vectors. Here, we report the results of numerical experiments for the cases of pure and mixed states considering the sources of randomness affecting the performance of the proposed method.

To study the violation of the CHSH inequality with an unknown state $\rho$, pure or mixed, we compute the expected value $\bar S(\rho)$ by sampling a sufficiently large number of independent trajectories, each obtained through the optimization of $S$ by CSPSA for $\rho$, as
\begin{equation}
\bar S(\rho)=\frac{1}{K}\sum_{{\bm z}_0,\{ {\bm\Delta}_1,\dots,{\bm\Delta}_k\}}S(\rho,{\bm z}_0,\{ {\bm\Delta}_1,\dots,{\bm\Delta}_k\}),
\end{equation}
where $S(\rho,{\bm z}_0,\{ {\bm\Delta}_1,\dots,{\bm\Delta}_k\})$ is the value of the CHSH function evaluated on a particular trajectory generated by a single run of CSPSA and $K$ is the total number of simulated trajectories. $S(\rho,{\bm z}_0,\{ {\bm\Delta}_1,\dots,{\bm\Delta}_k\})$ depends on the unknown state $\rho$, the set ${\bm z}_0$ of complex numbers that defines the initial guess for the four local measurements, and the particular sequence of choices $\{ {\bm\Delta}_1,\dots,{\bm\Delta}_k\}$. The mean $\bar S(\rho)$ will be studied as a function of the number $k$ of iterations for a fixed ensemble size $N$.

Since we are interested in the overall behavior of the algorithm for unknown states, we calculate the mean $\bar S_C$ of $\bar S(\rho)$ in a set $\Omega_C$, that is,
\begin{equation}
\bar S_C=\frac{1}{M}\sum_{\rho\in\Omega_C} \bar S(\rho),
\end{equation}
where $M$ is the number of states in $\Omega_C$ and $C$ is a parameter that characterizes the states in the set. Alternatively, we calculate the median $\tilde S_C$ of $\bar S(\rho)$ in the set $\Omega_C$ and the interquartile range. This is done to determine whether the distribution of $\bar S(\rho)$ in $\Omega_C$ exhibits a symmetric distribution or not and the possible existence of outliers.

\begin{figure}[!t]
	\centering
	\includegraphics[width=0.45\textwidth]{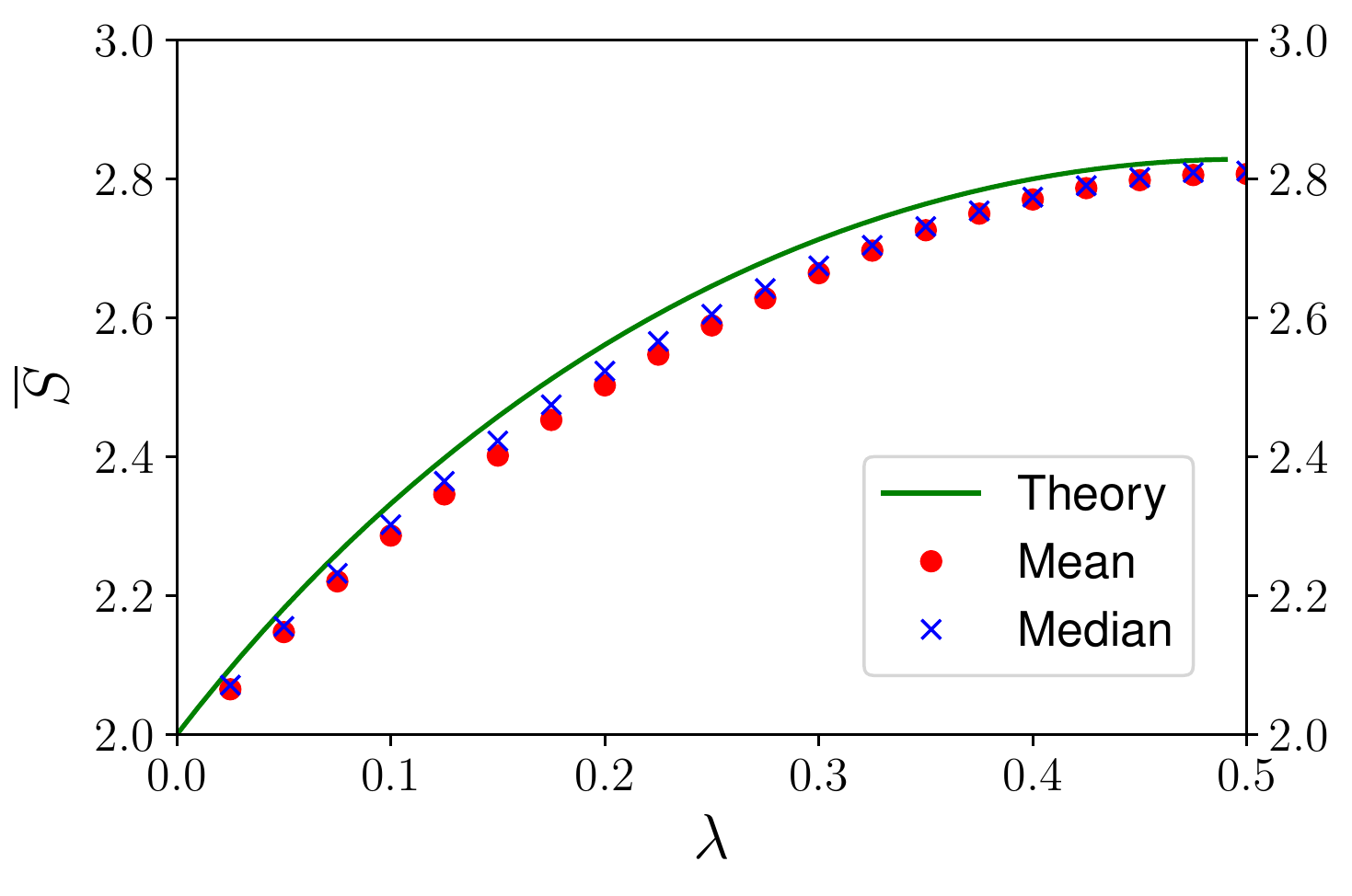}
	\caption{CHSH function $S(|\psi_\lambda\rangle)$ as a function of the Schmidt coefficient $\lambda$ for two-qubit states with fixed local Schmidt bases. Continuos green line represents the theoretical prediction given by Eq.~(\ref{CHSHfixedSchmidtbases}). Solid red circles (blue x's) represent the mean $\bar S(|\psi_\lambda\rangle)$ (median $\tilde S(|\psi_\lambda\rangle)$) of $S(|\psi_\lambda\rangle)$ obtained via CSPSA considering $10^4$ initial guesses for each state $|\psi_\lambda\rangle$, 200 iterations, and an ensemble size $N=10^2$.}
\label{Figure0a}
\end{figure}

\begin{figure}[!t]
	\centering
	\includegraphics[width=0.45\textwidth]{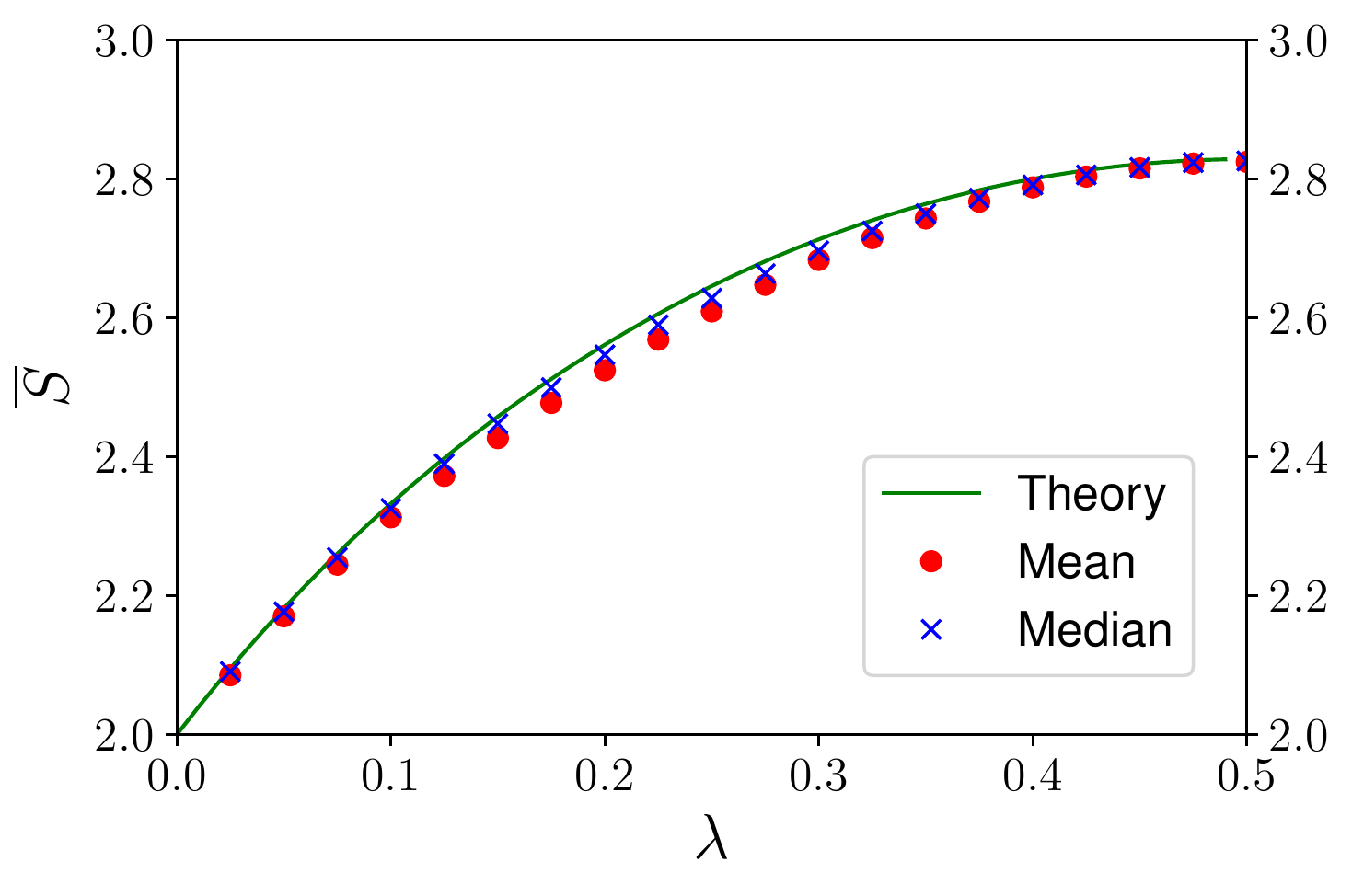}
	\caption{CHSH function $S(|\psi_\lambda\rangle)$ as a function of the Schmidt coefficient $\lambda$ for two-qubit states with fixed local Schmidt bases. Continuos green line represents the theoretical prediction given by Eq.~(\ref{CHSHfixedSchmidtbases}). Solid red circles (blue x's) represent the mean $\bar S(|\psi_\lambda\rangle)$ (median $\tilde S(|\psi_\lambda\rangle)$) of $S(|\psi_\lambda\rangle)$ obtained via CSPSA considering $10^4$ initial guesses for each state $|\psi_\lambda\rangle$, 200 iterations, and an ensemble size $N=10^4$.}
\label{Figure0b}
\end{figure}

\subsection{Unknown pure states}

We start our analysis of the proposed algorithm by considering the violation of the CHSH inequality for the set $\Omega_{\lambda}$ of two-qubit pure states defined by the Schmidt decomposition
\begin{equation}
|\psi(\lambda)\rangle=\sqrt{\lambda}|0\rangle_1|0\rangle_2+\sqrt{1-\lambda}|1\rangle_1|1\rangle_2,
\end{equation}
where $\lambda\in[0,1/2]$ is the Schmidt coefficient and $\{|0\rangle_1,|1\rangle_1\}$ and $\{|0\rangle_2,|1\rangle_2\}$ are fixed local Schmidt bases of each qubit. States in $\Omega_\lambda$ lead to a value of the function $S$ given by
\begin{equation}
S(\lambda)=2\sqrt{1+4\lambda(1-\lambda)}.
\label{CHSHfixedSchmidtbases}
\end{equation}
In Fig. \ref{Figure0a} we show $\bar S(\rho_{\lambda})$ for $\rho_{\lambda}=|\psi_\lambda\rangle\langle\psi_\lambda|$ as a function of $\lambda$ for $N=10^2$ after 200 iterations and $K=10^4$. Initial guesses for the set of four local observables are randomly chosen. In particular, information about the fixed bases in $|\psi_\lambda\rangle$ has not been used to improve the performance of CSPSA. As is apparent from Fig.~\ref{Figure0a}, CSPSA provides mean and median of $S(|\psi_\lambda\rangle)$ that closely resemble the theoretical prediction of Eq.~(\ref{CHSHfixedSchmidtbases}) for any value of $\lambda$. A much better agreement can be obtained by increasing the ensemble from $N=10^2$ to $N=10^4$, which is illustrated in Fig.~\ref{Figure0b}.

\begin{figure}[!t]
	\centering
	\includegraphics[width=0.45\textwidth]{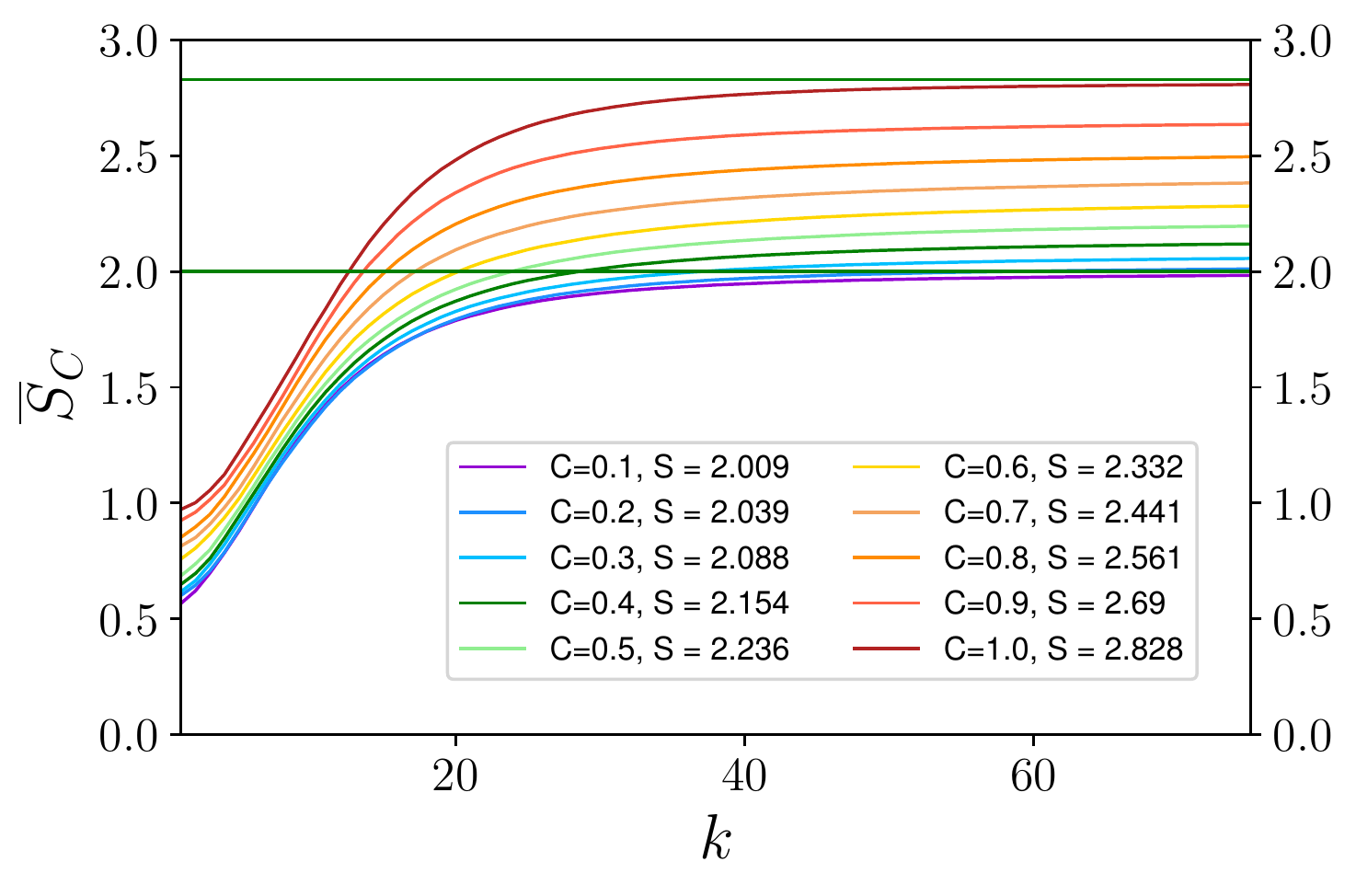}
	\caption{Mean $\bar S_C$ of $\bar S(\rho)$ in $\Omega_C$ as a function of the number of iterations for several values of the concurrence $C$ in the interval $[0.1, 1.0]$, from bottom to top. The mean $\bar S(\rho)$ is calculated with $10^4$ independent trajectories and each local measurement is simulated with an ensemble size $N=10^2$. Upper and lower straight lines represent the values 2$\sqrt{2}$ and 2, correspondingly.}
\label{Figure1}
\end{figure}

\begin{figure}[!t]
	\centering
	\includegraphics[width=0.45\textwidth]{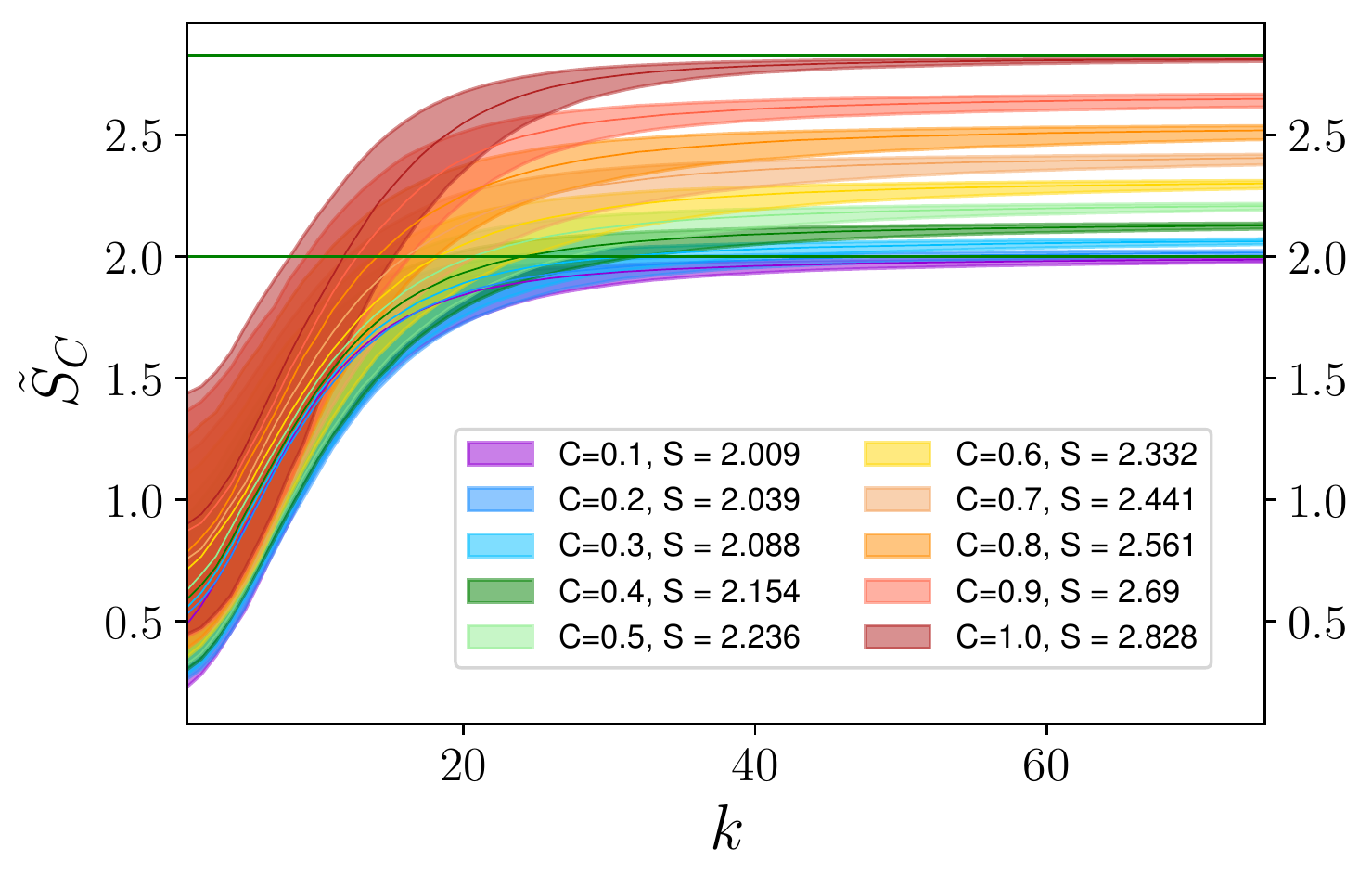}
	\caption{Median of $\bar S(\rho)$ in $\Omega_C$ as a function of the number of iterations for several values of the concurrence $C$ in the interval $[0.1, 1.0]$, from bottom to top. The mean $\bar S(\rho)$ is calculated with $10^4$ independent trajectories and each local measurement is simulated with an ensemble size $N=10^2$. Upper and lower straight lines represent the values $2\sqrt{2}$ and 2, correspondingly.}
\label{Figure2}
\end{figure}

Next we analyze the case of pure states with a known value of the concurrence $C$, which is given by the expression
\begin{equation}
C(\lambda)=2\sqrt{\lambda}\sqrt{1-\lambda}.
\label{CONCURRENCE}
\end{equation}
The local Schmidt bases of the state are unknown. In the simulations we choose a fixed value $C$ of the concurrence, which in turn fixes the value of the Schmidt coefficient. The local Schmidt bases are randomly chosen. As in the previous simulations, the knowledge about the value of the concurrence is not employed to improve the performance of CSPSA. Figure\thinspace\ref{Figure1} shows the behavior of $\bar S_C$, which is the mean of $\bar S(\rho)$ calculated on a set $\Omega_C$ of pure states with a fixed value $C$ of the concurrence, as a function of the number $k$ of iterations for several values of $C$. Each set $\Omega_C$ contains 100 states chosen according to a Haar-uniform distribution and $\bar S(\rho)$ is calculated with $10^4$ trajectories. Each one of the four local measurements is simulated considering an ensemble size of $N=10^2$. According to Fig.\thinspace\ref{Figure1}, the quantity $\bar S_C$ exhibits a fast increase of the value of the CHSH function within the first tens of iterations followed by a linear behavior, which asymptotically approaches the maximal value of the function $S$ for the value $C$ of the concurrence. The overall behavior of $\bar S_C$ does not depend on the value of $C$. 

Figure~\ref{Figure2} displays the median $\tilde S_C$ of $\bar S(\rho)$ in $\Omega_C$ as a function of the number of iterations for several values of the concurrence $C$. Shaded areas represent the interquartile range. Monte Carlo experiments are carried out as in Fig.~\ref{Figure1}. As is apparent from this figure, the median exhibits the same overall behavior as the mean $\bar S_C$. Mean and median reach after a few tens interations values that are nearly indistinguishable and contained within the interquartile range. This indicates that the stochasticity of CSPSA does not lead to outliers in the histogram of $\bar S(\rho)$ for all simulated sets $\Omega_C$. The interquartile range, which is a quartile-based measure of variability, decreases rapidly with the number of iterations and becomes a very narrow fringe. This is an indication that the histogram of $\bar S(\rho)$ for a particular $\Omega_C$ after a few tens iterations is highly concentrated around the mean. 

Thus, Figs.\thinspace\ref{Figure1} and \ref{Figure2} clearly indicate that CSPSA can be employed to iteratively increase the value of the CHSH function for unknown pure states and detect entanglement. The greater the entanglement of the unknown state, the fewer iterations will be required to obtain a violation of the CHSH inequality. Furthermore, approximately 70 iterations are necessary to reach a value of the CHSH function close to the maximal violation allowed by quantum mechanics.

\begin{figure}[t]
	\centering
	\includegraphics[width=0.45\textwidth]{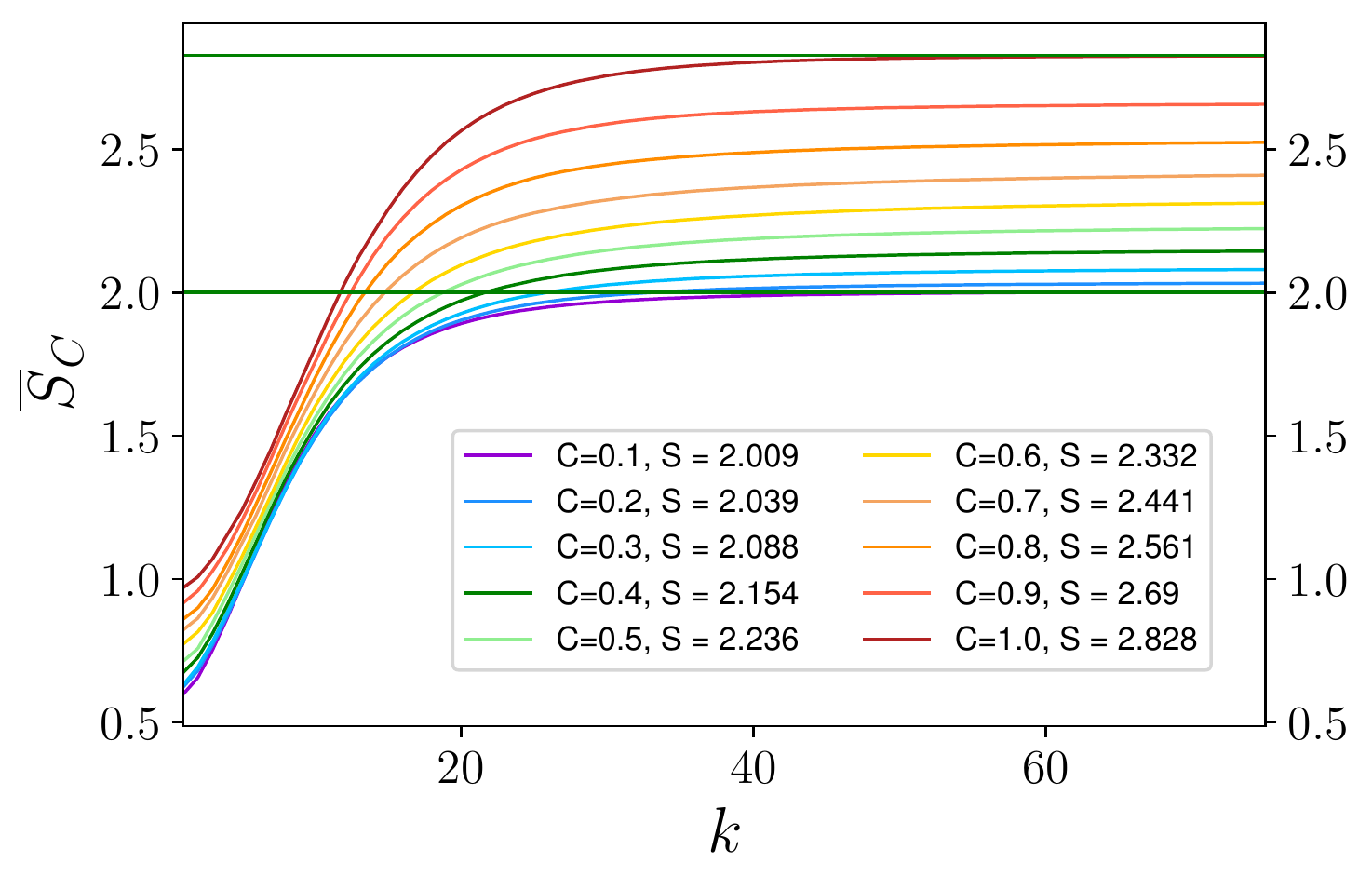}
	\caption{Mean $\bar S_C$ of $\bar S(\rho)$ in $\Omega_C$ as a function of the number of iterations for several values of the concurrence $C$ in the interval $[0.1, 1.0]$, from bottom to top. The mean $\bar S(\rho)$ is calculated with $10^4$ independent trajectories and each local measurement is simulated with an ensemble size $N=10^4$. Upper and lower straight lines represent the values $2\sqrt{2}$ and 2, correspondingly.}
\label{Figure3}
\end{figure}

\begin{figure}[!t]
	\centering
	\includegraphics[width=0.45\textwidth]{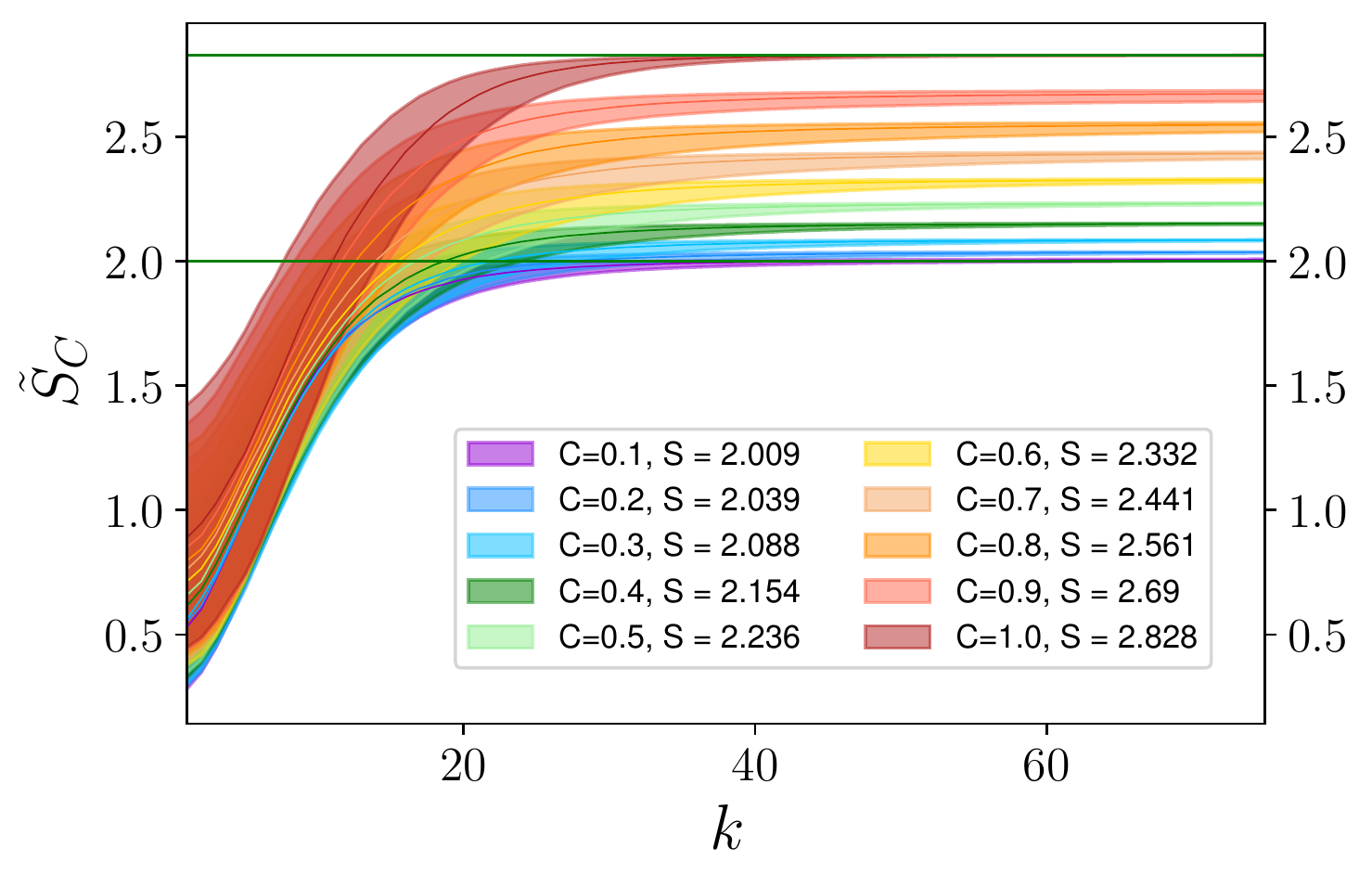}
	\caption{Median of $\bar S(\rho)$ in $\Omega_C$ as a function of the number of iterations for several values of the concurrence $C$ in the interval $[0.1, 1.0]$, from bottom to top. The mean $\bar S(\rho)$ is calculated with $10^4$ independent trajectories and each local measurement is simulated with an ensemble size $N=10^4$. Upper and lower straight lines represent the values $2\sqrt{2}$ and 2, correspondingly.}
\label{Figure4}
\end{figure}

Figs.\thinspace\ref{Figure3} and \ref{Figure4} depicts the mean $\bar S_C$ and the median $\tilde S_C$ of $\bar S(\rho)$ in $\Omega_C$, correspondingly. In this case local measurements are simulated with an ensemble size of $N=10^4$, that is, a quadratic increase with respect to previous simulations. As is apparent from Figs.\thinspace\ref{Figure3} and \ref{Figure4}, the overall behavior remains unchanged with respect to Figs.~\ref{Figure1} and \ref{Figure2}. In particular, both values of ensemble size, $N=10^2$ and $N=10^4$, show small differences in the asymptotic linear regime. For instance, for weakly entangled states, that is, $C=0.1$, after the total of iterations, in Fig.~\ref{Figure2} CSPSA is close to 2 but below. In Fig.~\ref{Figure4}, CSPSA is slightly above 2. Similar differences can be observed for other values of $C$. Furthermore, a small reduction in the number of iterations required to violated the CHSH inequality can be observed. This reduction depends on the initial amount of entanglement of the unknown state. Also, the increase in $N$ leads to narrower interquartile ranges.

\begin{figure}[!t]
	\centering
	\includegraphics[width=0.45\textwidth]{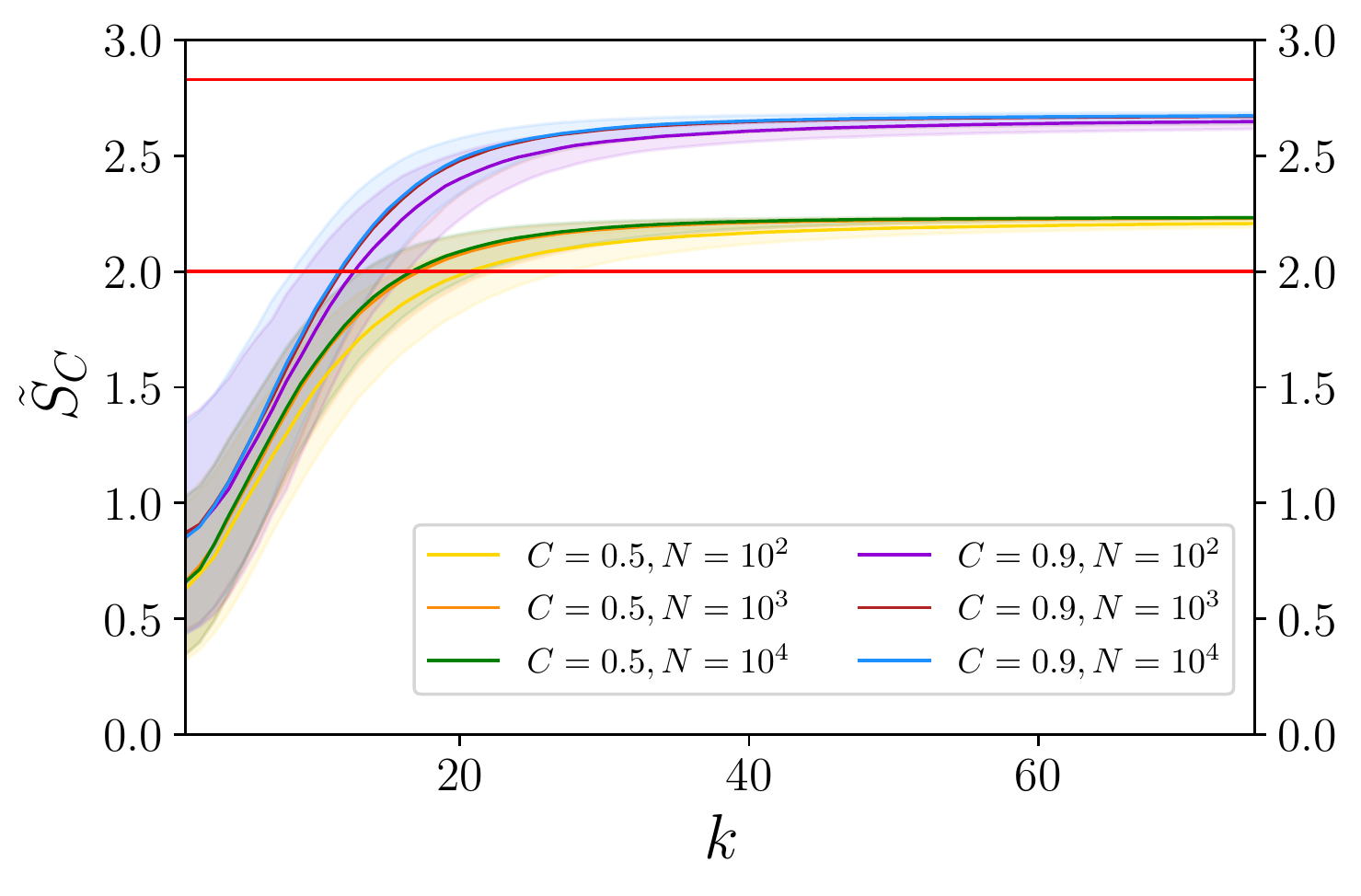}
	\caption{Median $\tilde S_C$ of $\bar S(\rho)$ in $\Omega_C$ as a function of the number of iterations for $C=0.5$ and $C=0.9$. Each local measurement is simulated with an ensemble size $N=10^4, 10^3, 10^2$. The median $\tilde S(\rho)$ for each value of $C$ is calculated with $10^4$ independent trajectories. Upper and lower straight lines represent the values $2\sqrt{2}$ and 2, correspondingly.}
\label{Figure5}
\end{figure}

This is more clearly illustrated in Fig.\thinspace\ref{Figure5}, which shows the median $\tilde S_C$ of $S(\rho)$ in $\Omega_C$ for $C=0.5$  and  $C=0.9$ for three values of ensemble size $N=10^2, 10^3, 10^4$. The interquartile range is also depicted. As is apparent from Fig.\thinspace\ref{Figure5}, CSPSA provides very similar values of $\tilde S_C$ almost independently of the size of the ensemble employed. However, in the regime of a few tens of iterations, $N=10^2$ leads to lower values of $\tilde S_C$, while $N=10^3$ and $10^4$ lead to very similar values of $\tilde S_C$, which are higher than in the case $N=10^2$. This has for consequence that higher values of $N$ lead to a decrease in the number of iterations required to observe a violation of the CHSH inequality, \lp{but this improvement is saturated for an enough large sample size.}

\begin{figure}[!t]
	\centering
	\includegraphics[width=0.45\textwidth]{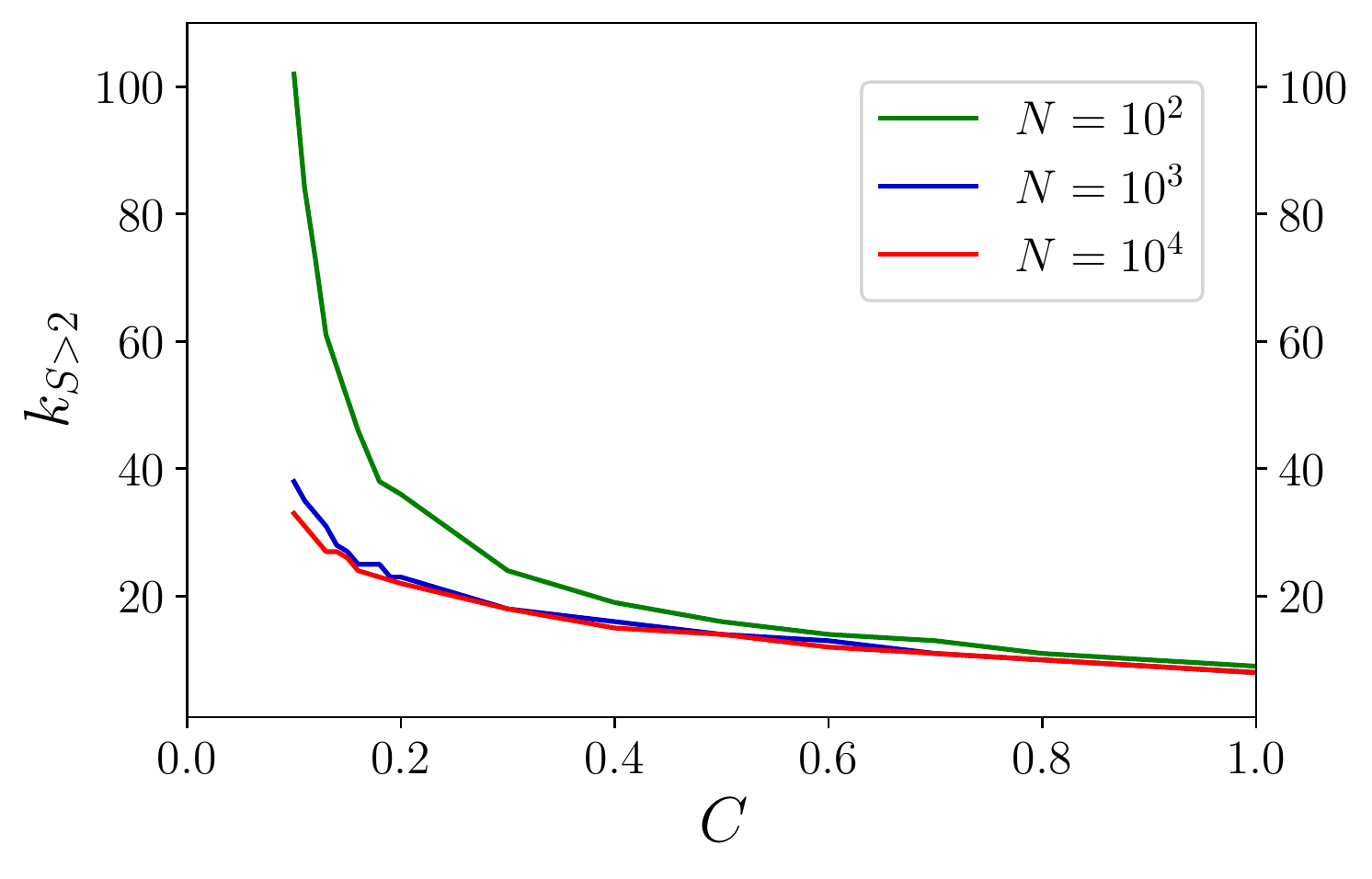}
	\caption{Number of iterations $k_{S>2}$ such that the interquartile range is above $S=2$ as a function of the concurrence $C$ for $N=10^2,10^3$, and $10^4$, from top to bottom.}
\label{Figure6}
\end{figure}

This later effect is analyzed with the help of Fig.\thinspace\ref{Figure6} that displays the number of iterations $k_{S>2}$ required to obtain a violation of the inequality with $75\%$ of the states generated for a given value of $C$ and with $N=10^2, 10^3, 10^4$. Here, we observe that $N=10^4$ and $N=10^3$ lead to a very similar behavior while $N=10^2$ requires the largest number of iterations to reach a violation  of the CHSH inequality. Also, the lower the concurrency value, the greater the number of iterations required for the violation. In fact, Fig.\thinspace\ref{Figure6} suggests that $k_{S>2}$ decreases exponentially with $C$. This figure also illustrates the interplay between  $k_{S>2}$ and the total ensemble size $N_{S>2}$ required for violating the CHSH inequality. For example, in the case of $C=0.1$ and $N=10^2$,  we have that approximately $k_{S>2}=100$, which leads to $N_{S>2}=8\times10^4$. For $N=10^4$ we have that approximately $k_{S>2}=35$ and thus $N_{S>2}=280\times10^4$. Clearly, the reduction in the value of $k_{S>2}$ comes at the expense of using a much larger total ensemble $N_{S>2}$. For states with a high value of concurrence $C$, the reduction in the value of $k_{S>2}$ by increasing the value of $N$ is marginal. 

\begin{figure}[!t]
	\centering
	\includegraphics[width=0.45\textwidth]{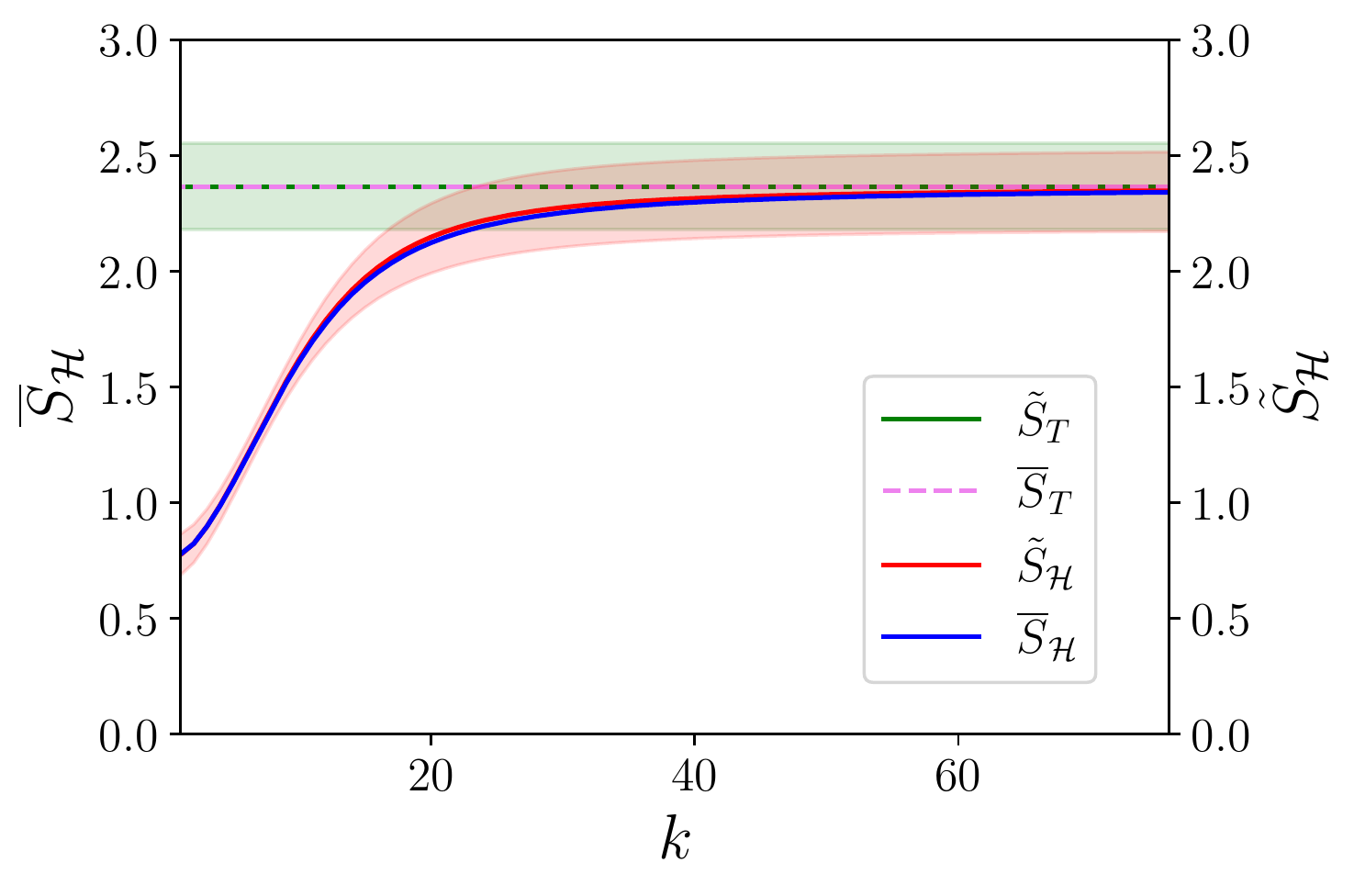}
	\caption{Mean $\bar S_{\cal H}$ and median $\tilde S_{\cal H}$ of $\bar S(|\psi\rangle\langle\psi|)$ with $|\psi\rangle\in\cal H$ and interquartile range for $N=10^2$.}
\label{Figure7}
\end{figure}

\begin{figure}[!t]
	\centering
	\includegraphics[width=0.45\textwidth]{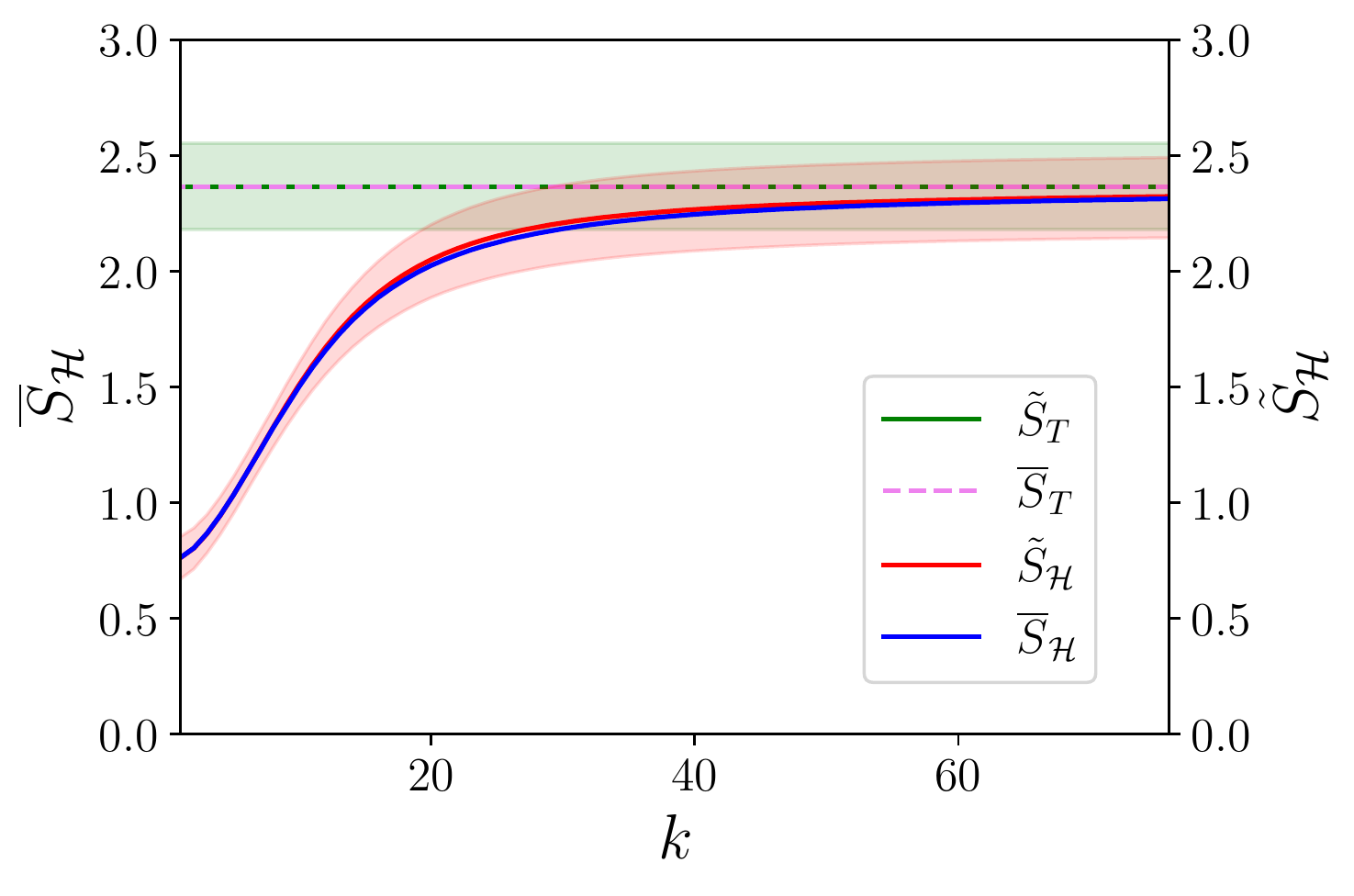}
	\caption{Mean $\bar S_{\cal H}$ and median $\tilde S_{\cal H}$ of $\bar S(|\psi\rangle\langle\psi|)$ with $|\psi\rangle\in\cal H$ and interquartile range for $N=10^4$.}
\label{Figure8}
\end{figure}

So far, our study of the violation of CHSH inequality through CSPSA has been done considering that the initial amount of entanglement is known. This was done to show that CSPSA drives the value of the CHSH function $S$ close to the maximum value regardless of the amount of entanglement. We now lift this assumption and consider unknown pure states. In order to do this, we generate a set $\Omega_{\cal H}$ with 100 pure states in the Hilbert space ${\cal H}={\cal H}_1\otimes{\cal H}_2$ of two qubits according to a Haar-uniform distribution and calculate the mean $\bar S_{\cal H}$ and the median $\tilde S_{\cal H}$ of $\bar S(|\psi\rangle\langle\psi|)$ in $\Omega_{\cal H}$, together with the corresponding interquartile range. These quantities are depicted in Fig.~\ref{Figure7} as a function of the number of iterations. The behavior exhibited by the mean and media is very similar and characterized by a fast increase within the first tens of iterations followed by an asymptotic linear regime. Fig.~\ref{Figure7} also shows the mean and media of the maximal theoretical values of $S$ for each state in $\Omega_{\cal H}$, which are indicated as two superposed straight lines. As can be seen from Fig.~\ref{Figure7}, CSPSA produces a mean and a median that are very closely to the theoretical values. Also, the expected number of iterations $k_{S>2}$ such that 75\% of the simulated states violates the CHSH inequality is about 25. Fig.~\ref{Figure8} shows the same information as Fig.~\ref{Figure7} but with $N=10^4$. In this case, we see that the quadratic increase in the ensemble size allows CSPSA to reach mean and media values that are even closer to the theoretical values. Furthermore, there is a small reduction in the number of iterations required to obtain a value of $S$ greater than two from 25 to 20.

Our previous simulations seem to indicate that the optimization of the CHSH function for an unknown state through the CSPSA method provides maximum values of the CHSH functional close to the theoretical maximum values. In order to analyze this we employ the mean square error. For a given state $\rho=|\psi\rangle\langle\psi|$ and a single realization of CSPSA we calculate the square error $SE(\rho)$ as
\begin{equation}
SE(\rho)=|S(\rho,{\bm z}_0,\{ {\bm\Delta}_1,\dots,{\bm\Delta}_k\}-S_{max}(\rho)|^2.
\end{equation}
The mean square error $MSE(\rho)$ for a fixed unknown state $\rho$ with respect to a large set of realizations is given by
\begin{equation}
MSE(\rho)=\frac{1}{K}\sum_{{\bm z}_0,\{ {\bm\Delta}_1,\dots,{\bm\Delta}_k\}}SE(\rho),
\end{equation}
which corresponds to an estimation accuracy metric. This is then used to calculate the average of the mean square error $\overline{MSE}$ on the total Hilbert space $\cal H$ as
\begin{equation}
\overline{MSE}=\frac{1}{M}\sum_{\rho\in\Omega_{\cal H}} MSE(\rho).
\end{equation}

\begin{figure}[!t]
	\centering
	\includegraphics[width=0.45\textwidth]{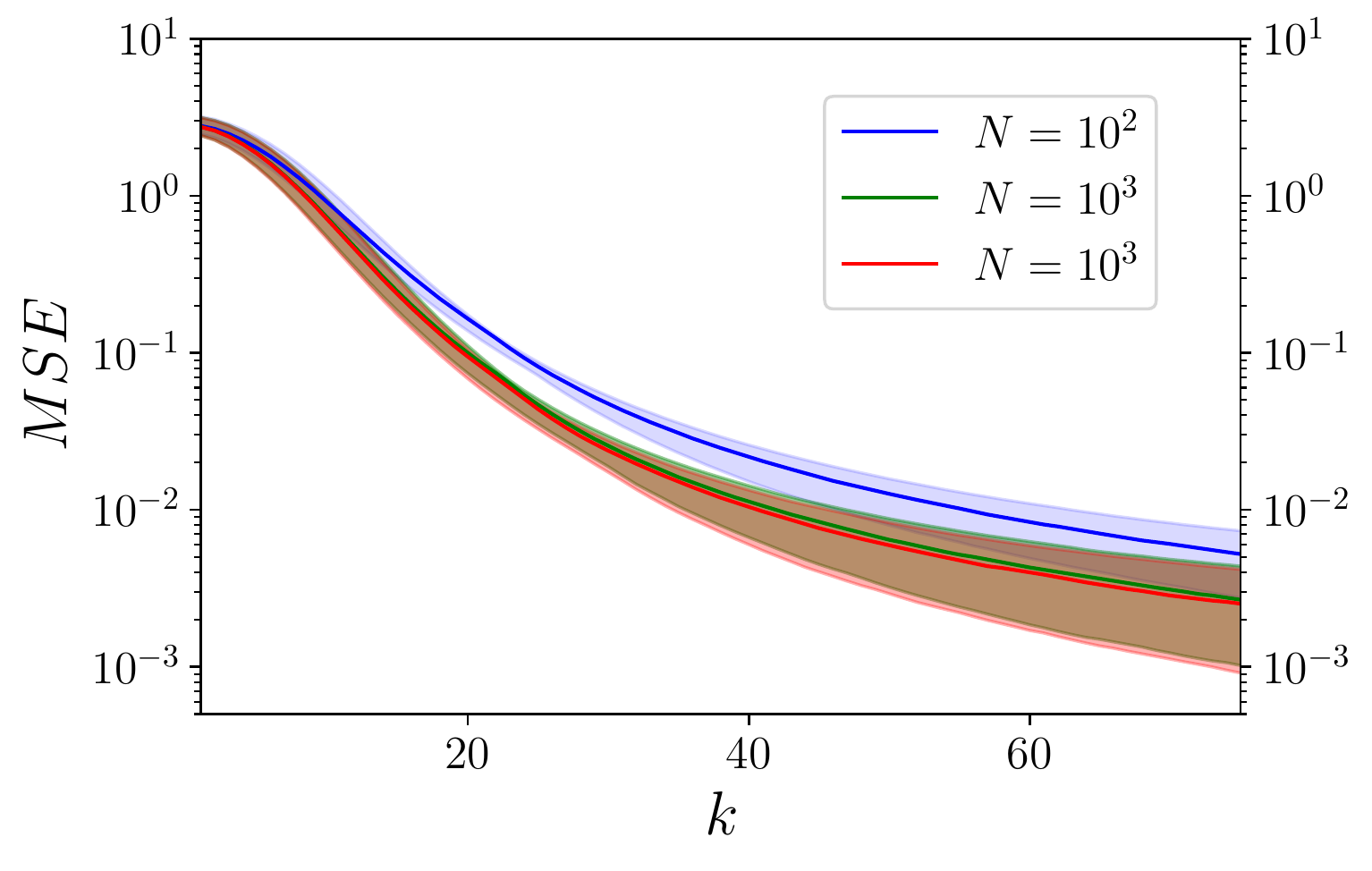}
	\caption{Mean square error $\overline{MSE}$ as a function of the number $k$ of iterations for $N=10^2, 10^3$, and $10^4$, from top to bottom. Shaded areas represent interquartile range.}
\label{Figure9}
\end{figure}

Figure~\ref{Figure9} shows the mean $\overline{MSE}$ of the square error on the Hilbert space as a function of the number of iterations for $N=10^2, 10^3, 10^4$. For each value of ensemble size, $\overline{MSE}$ displays a fast decrease followed by an approximately asymptotic lineal behavior. $N=10^3$ and $N=10^4$ produce very similar values of the mean square error while an ensemble size of $N^2$ produces a value that is almost half order of magnitude higher. After 25 iterations the difference between the maximal theoretical value and the value achieved by CSPSA is between $10^{-1}$ and $10^{-2}$. Adding 50 more iterations this difference is approximately between $10^{-2}$ and $10^{-3}$. Let us recall that after 75 iterations the lower bound of the interquartile range of $\bar S(|\psi\rangle\langle\psi|)$ has an approximate value of $2.12$, so that for 75\% of states in the bipartite Hilbert space we can ascertain its entangled nature and assign an accurate value of the CHSH function. A further improvement in the accuracy achieved by CSPSA can be obtained at the expense of a large increase in the number of iterations, after adding 150 iterations we obtain a new decrease by one order of magnitude, that is, the mean $\overline{MSE}$ of the square error on the Hilbert space is approximately in the interval between $10^{-3}$ and $10^{-4}$.

\subsection{Unknown mixed states}

In the previous section, we have studied the violation of the CHSH inequality for unknown pure states by means of a CSPSA-driven sequence of local measurements. Here, we study the case of mixed bipartite states.

We start by reproducing the value of the CHSH function on the set of the Werner states, which are given by the expression
\begin{equation}
\rho_\lambda=\lambda|\psi_s\rangle\langle\psi_s|+\frac{(1-\lambda)}{d}\emph{I},
\end{equation}
where $|\psi_s\rangle$ is the the maximally entangled singlet state defined as
\begin{equation}
|\psi_s\rangle=\frac{1}{\sqrt{2}}(|0\rangle|1\rangle-|1\rangle|0\rangle)
\end{equation}
and $\emph{I}$ is a 4-dimensional identity operator. This mixture of the singlet state with white noise is separable if and only if $\lambda\le1/3$ and violates the CHSH inequality if and only if $\lambda>1/\sqrt{2}$. The maximal value of the CHSH function for a Werner state $\rho_\lambda$ is given by
\begin{equation}
S(\rho_\lambda)=2\sqrt{2}\lambda.
\label{CHSHWERNER}
\end{equation}

\begin{figure}[!t]
	\centering
	\includegraphics[width=0.45\textwidth]{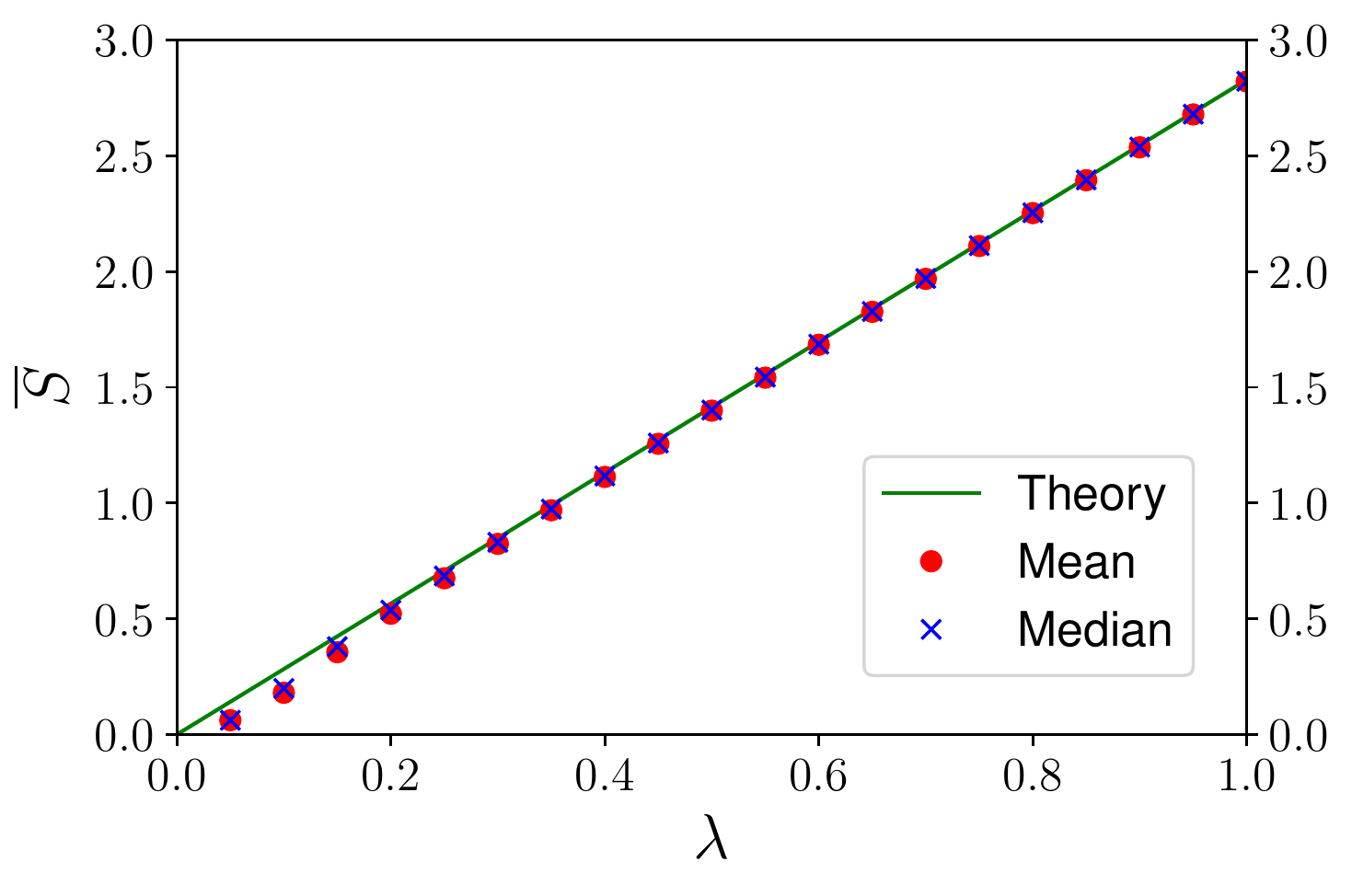}
	\caption{Mean $\bar S(\rho_\lambda)$ (solid red dots) and median $\tilde S(\rho_\lambda)$ (blue x's) as a function of $\lambda$ for Werner states. Continuous black line depicts the maximal value of the CHSH function of Eq.~(\ref{CHSHWERNER}). Local measurements are simulated with an ensemble size $N=10^2$ and 75 iterations are realized.}
\label{Figure10}
\end{figure}

\begin{figure}[!t]
	\centering
	\includegraphics[width=0.45\textwidth]{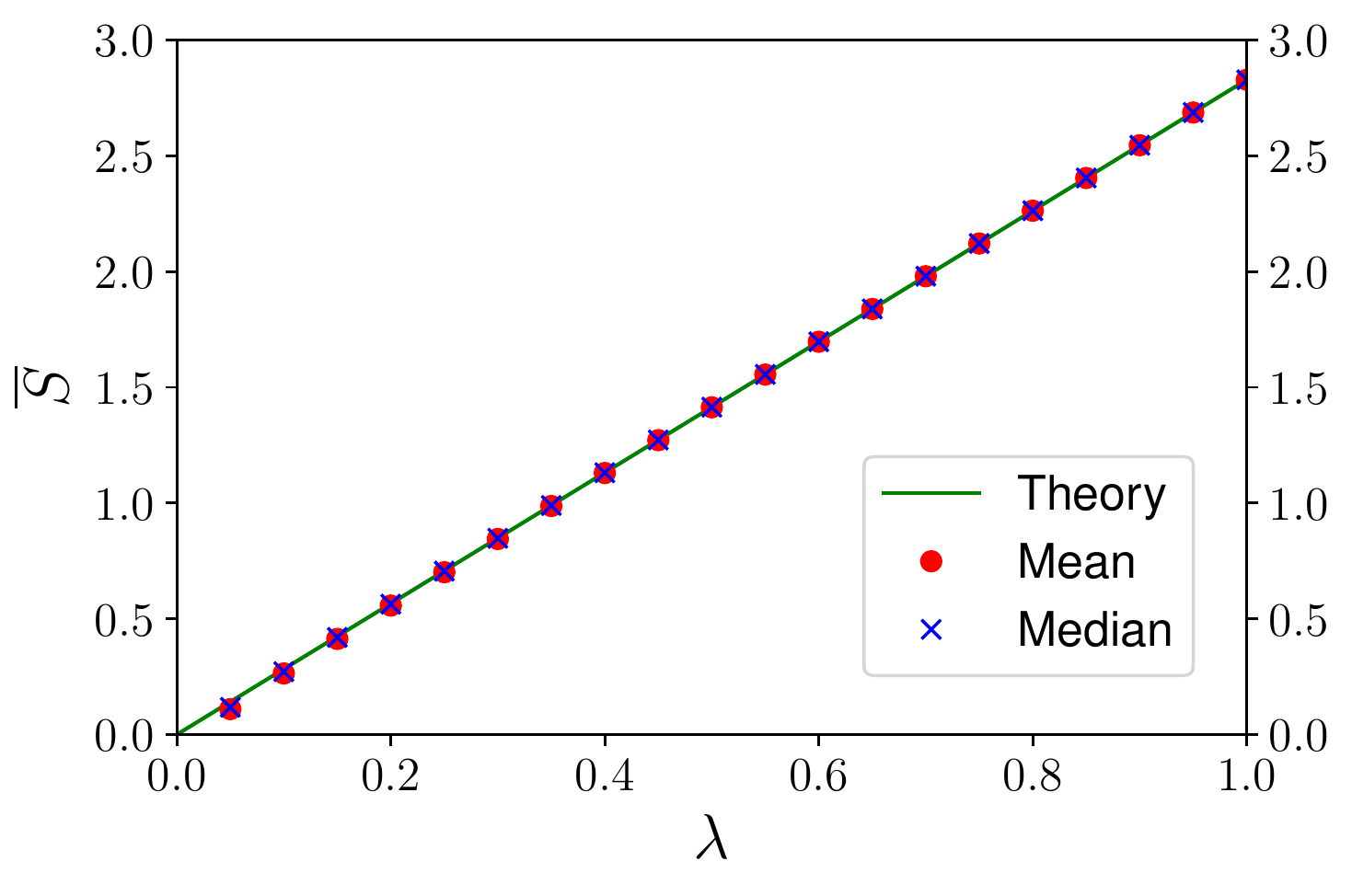}
	\caption{Mean $\bar S(\rho_\lambda)$ (solid red dots) and median $\tilde S(\rho_\lambda)$ (blue x's) as a function of $\lambda$ for Werner states. Continuous black line depicts the maximal value of the CHSH function of Eq.~(\ref{CHSHWERNER}). Local measurements are simulated with an ensemble size $N=10^4$ and 75 iterations are realized.}
\label{Figure11}
\end{figure}

Figure~\ref{Figure10} displays the mean $\bar S(\rho_\lambda)$ and median $\tilde S(\rho_\lambda)$ as a function of $\lambda$ obtained via CSPSA for an ensemble size $N=10^2$ after 75 iterations. With the exception of the first 5 points, Figure \ref{Figure10} shows a very good agreement between the maximal value of the CHSH function of Eq.~(\ref{CHSHWERNER}) and the value achieved with the help of CSPSA. Furthermore, mean and median exhibit values that also are very close and the interquartile (not depicted) range is very narrow. Thus, within the family of Werner states CSPSA drives the sequence of local measurement bases very close to the optimal set. An increase in the ensemble size leads to even better results. This is illustrated in Fig.~\ref{Figure11}, where local measurements are simulated with an ensemble $N=10^4$. In this case all points are closer to the maximal value of the CHSH inequality. 

Next we proceed with the case of unknown mixed states. We randomly generated a set of $10^6$ two-qubit mixed states. In order to determine whether a mixed state violates or not the CHSH inequality we employ the $M$ quantity criterion \cite{1}. A mixed state $\rho$ acting on a Hilbert space $\mathcal{H}=\mathcal{H}_2\otimes \mathcal{H}_2$ can be represented in the form
\begin{eqnarray}
\rho &=& \frac{1}{4}\left(I\otimes I +\sum_{i=1}^3r_i\sigma_i\otimes I + I\otimes \sum_{i=1}^3s_i\sigma_i\right.
\nonumber\\
&+& \left.\sum_{n,m=1}^3 t_{nm}\sigma_n\otimes\sigma_m \right), \label{rho}
\end{eqnarray}
where $I$ represents the 2-dimensional identity operator, $\{\sigma_n\}_{n=1}^3$ are the standard Pauli matrices, and the real coefficients $r_i, s_i$ and $t_{n,m}$ define the mixed state. The quantity $M$ is defined by $M(\rho)=u+\tilde{u}$, where $u$ and $\tilde{u}$ denote the greater positive eigenvalues of the matrix $U_{\rho}:=T_{\rho}^T T_{\rho}$ being the coefficients of the matrix $T(\rho)$ given by $t_{nm}=\mbox{Tr}(\rho\sigma_n\otimes\sigma_m)$. A state $\rho$ violates the CHSH inequality if and only if 
the condition $M(\rho)~>~1$ holds \cite{1}. Employing this criterium, the initial set of $10^6$ mixed states was reduced to a set $\Omega$ containing $8\times10^3$ mixed states with $M(\rho)>1$ that violate the CHSH inequality.

To analyze the values of the CHSH function obtained through CSPSA we use those obtained through SDP. In the SDP case we need to fix the state that is used in the maximization. However, let us recall that even when the states are fixed, the maximization of $S$ remains to be a nonlinear problem. Therefore, to find the maximum value of $S$ for each state in $\Omega$ we use the see-saw method \cite{2,3} to iterate a SDP test \cite{4,5} where either observable A or B remain fixed while optimizing in the other variable. 
The SDP that we solve is the following
\begin{eqnarray}
\mbox{\rm given}~&& \rho_\Omega, A(z_a), A(z_a'), \\
\underset{B(z_b), B(z_b')}{\mbox{max}} && S(\rho_\Omega,A(z_a), A(z_a'),B(z_b), B(z_b')),
\end{eqnarray}
with the conditions
\begin{eqnarray}
|\Psi(z_b)\rangle\langle\Psi(z_b)|,|\Psi^{\perp}(z_b)\rangle\langle\Psi^{\perp}(z_b)|\geq 0 \quad  \forall \ z_b,z_b',\\
|\Psi(z_b)\rangle\langle\Psi(z_b)| + |\Psi^{\perp}(z_b)\rangle\langle\Psi^{\perp}(z_b)| = I \quad \forall \ z_b,z_b'.
\end{eqnarray}
Notice that this SDP takes Alice's observables $A(z_a)$ and $A(z_a')$ as inputs and for a given mixed state from the $\Omega$ set, it finds Bob's observables $B(z_b)$ and $B(z_b')$ that maximally violate $S$. Then, we take the observables $B$ outputted by this SDP as inputs in a new iteration to obtain optimal observables $A$. This procedure is iterated until some suitable convergence condition is satisfied. We performed this optimization for every mixed bipartite state in the set $\Omega$, which allows us to find better lower bounds on $S$, together with the optimal observables $A$ and $B$. 

\begin{figure}[!t]
	\centering
	\includegraphics[width=0.45\textwidth]{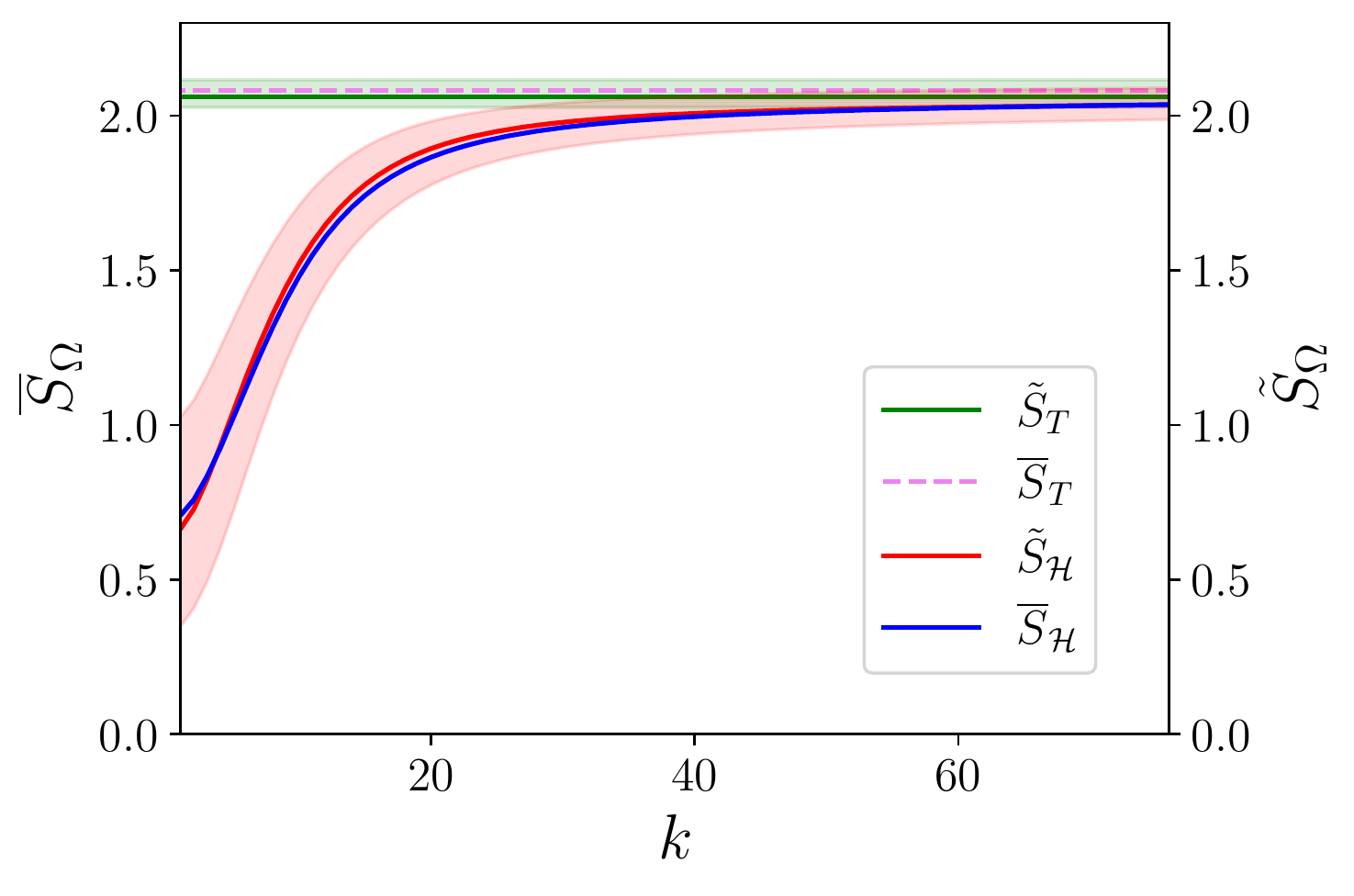}
	\caption{Mean $\bar S_{\Omega}$ (red solid line) and median $\tilde S_{\Omega}$ (blue solid line) obtained via CSPSA on the set $\Omega$ of randomly generated mixed entangled states as a function of the number $k$ of iterations. Mean $\bar S_{\Omega}$ (yellow solid line) and median $\tilde S_{\Omega}$ (green solid line) obtained via SDP on the set $\Omega$ of randomly generated mixed entangled states as a function of the number $k$ of iterations. Shaded areas correspond to interquartile range. CSPSA simulations consider ensemble size $N=10^4$.}
\label{Figure12}
\end{figure}

Figure~\ref{Figure12} displays the behavior of the mean $\bar S_{\Omega}$, median $\tilde S_{\Omega}$, and interquartile range as functions of the number of iterations. This figure also displays the values of these quantities obtained via SDP. As is apparent from this figure, the values of the mean and median provided via CSPSA are very close and tend to agree with the values delivered by SDP after tens of iterations. Also, the interquartile ranges tend to overlap. However, in the case of mixed states the number of iterations needed to obtain a violation of the CHSH inequality is much greater than in the case of pure states. This is due to the fact that the mixed states in $\Omega$ typically have small values of the negativity, a well-known measure of entanglement, and thus, as in the case of weakly-entangled pure states, need more iterations to reach a violation of the CHSH inequality.

\section{Conclusions}

We have studied the problem of detecting the entanglement of unknown two-qubit states, mixed or pure, by violating the Clauser-Horne-Shimony-Holt inequality. Our approach to this problem is based on the maximization of the CHSH function by means of a stochastic optimization method, the Complex simultaneous perturbation stochastic approximation. This allows optimizing functions with unknown parameters, which in our case correspond to the unknown quantum state. CSPSA employs an iterative rule which requires at each iteration the value of the target function, that is, the CHSH function, at two different points in the optimization space. This is formed by vectors on the field of the complex numbers containing the measurement settings of four observables. The values of the CHSH function can be experimentally obtained even if the two-qubit state remains unknown. Thereby, CSPSA generates a sequence of measurement settings that in mean lead to increasing values of the CHSH inequality.

To analyze the characteristics of the proposed method, we carried out several numerical experiments. In particular, due to the stochastic nature of CSPSA, we employ random sampling to obtain estimates of the mean, median, and interquartile range of the quantities of interest. We first note that for a fixed unknown state, CSPSA provides very similar values of the mean and median of the CHSH function and a very narrow interquartile range. This indicates that CSPSA does not generates outliers, that is, for a given unknown state different realizations of our method provide very close results. This feature has been observed for each state in a universe of $5\times10^4$ randomly generated pure two-qubit states. 

The typical behavior of the mean of the CHSH function, as a function of the number of iterations, corresponds to a rapid increase followed by an approximately linear asymptotic behavior, which approaches the maximal value of the CHSH function. Unknown states characterized by the same concurrence value exhibit a very similar behavior of the CHSH function. However, the rate of convergence towards the maximum depends on the initial value of the concurrence. The higher the concurrence value, the fewer iterations are required to obtain a violation of the CHSH inequality and, consequently, detect entanglement. For example, states with maximum concurrence need 13 iterations while states with a concurrence of 0.1 need approximately 75 iterations to reach a violation. The number of iterations required to detect entanglement can be decreased by increasing the size of the ensemble of identically prepared copies that is employed to estimate the expectation values entering in the CHSH function. In our simulations, however, the effect of increasing the ensemble size is more notorious in the case of highly entangled states. We have studied the mean of the CHSH function on the 2-qubit Hilbert space. In this case, for an ensemble size of $10^2$ the entanglement of the randomly generated states is detected in mean by violating the CHSH inequality after 17 iterations, while after 25 iterations 75\% of the randomly generated states violate the CHSH inequality. These figures can be reduced by increasing the ensemble size. We have also studied the accuracy provided by our method in the estimation of the maximum value of the CHSH function. As accuracy metric we have used the mean squared error, which shows that after 25 iterations the difference between the maximal theoretical value and the value achieved by CSPSA is between $10^{-1}$ and $10^{-2}$. After 75 iterations the accuracy is approximately between $10^{-2}$ and $10^{-3}$. We have also considered the case of mixed states. The proposed method is capable of reproducing the maximal value of the CHSH function for Werner states and for randomly chosen mixed states. 

Therefore, the numerical simulations indicate that the maximization of the CHSH function through CSPSA leads to the detection of the entanglement of unknown states, pure or mixed. In mean, 25 iterations detect the entanglement of 75\% of the generated states. Also, it is possible to reach an accurate value of the maximal violation. 

There are some variations of the method here proposed that could reduce the number of iterations used to detect entanglement. We implement CSPSA considering the standard choice for the gain coefficients. However, these can be optimized. This is in general a difficult problem. Nevertheless, some simple heuristic prescriptions have been discussed in the study of various proposals of variational quantum eigensolvers \cite{Kandala}. These are based on SPSA, a version of CSPSA that works on the field of the real numbers. It seems possible that the SPSA performance-enhancing prescriptions could also be used to improve the CSPSA convergence rate, which would reduce the number of iterations required to detect entanglement. \lp{The usage of second-order methods or quantum natural gradient could also speed up the protocol \cite{Spall_2O, Wang_SPSA, Stokes_QN, Gacon_QN, Gidi_2O}. These employ additional measurements of the objective function to estimate its Hessian matrix, or fidelity to estimate the metric tensor. Thereafter, these matrices are used to precondition the gradient in order to improve the convergence rate, avoiding the need for tuning of some gain coefficients.} Another possibility arises when considering the large amount of information generated by our method. At each iteration 4 local observables are measured, which after several iterations provide a considerable amount of information about the unknown state. Thus, we can obtain an estimate of the unknown states by means of maximum likelihood \cite{Zambrano}. This, together with the estimate of the optimal measurement settings provided by CSPSA, can be used as initial guesses in a SDP problem to optimize the CHSH function. The solution of this problem can be used as the initial guess of the optimal measurement settings in the next iteration of CSPSA. This procedure does not increases the amount of measurements to be carried out but the computational cost. Besides, the use of {\it a priori} information can be employed to further increase the CSPSA convergence rate and achieve entanglement detection with a reduced number of iterations. 

\lp{We would like to remark that our approach based on CSPSA can be employed in other interesting problems. The construction of entanglement witnesses is a demanding computational task \cite{Zhu,Dai}, especially if the state is unknown, but it could be done efficiently with our method. The search for the optimal measurement settings to violate a multiqubit Bell inequality is challenging \cite{Svetlichny,Collins,Seevinck}. This is because the dimension scales exponentially with the number of qubits, so finding the optimal with quantum tomography and SPD is unfeasible. Our approach could provide an advantage in this problem since its resource scales with the number of iterations and not with the number of qubits.}

\section*{Acknowledgments}
This work was supported by ANID -- Millennium Science Initiative Program -- ICN17$_-$012 and by Fondo Nacional de Desarrollo Cient\'ifico y Tecnol\'ogico (FONDECYT) Grant No 1180558. J.\thinspace C.-V. was supported by CONICYT- PCHA/DoctoradoNacional/2018-21181692.  J. F. B. acknowledges support from FONDECYT Grant No 317030. L.P. was supported by ANID-PFCHA/DOCTORADO-BECAS-CHILE/2019-772200275, the CSIC Interdisciplinary Thematic Platform (PTI+) on Quantum Technologies (PTI-QTEP+), the CAM/FEDER Project No. S2018/TCS-4342 (QUITEMAD-CM), and the Proyecto Sinérgico CAM 2020 Y2020/TCS-6545 (NanoQuCo-CM).

\section*{Declarations}

The code and the data simulated to generate the figures are available on reasonable request.

\end{document}